\documentclass[english,jcp,superscriptaddress,preprint]{revtex4-2}
\usepackage[T1]{fontenc}
\usepackage[latin9]{inputenc}
\usepackage{color}
\usepackage{array}
\usepackage{textcomp}
\usepackage{multirow}
\usepackage{amstext}
\usepackage{amssymb}
\usepackage{graphicx}

\makeatletter

\newcommand{\lyxmathsym}[1]{\ifmmode\begingroup\def\b@ld{bold}
  \text{\ifx\math@version\b@ld\bfseries\fi#1}\endgroup\else#1\fi}

\providecommand{\tabularnewline}{\\}

\makeatother

\usepackage{babel}
\begin{document}
\title{Defect-driven tunable electronic and optical properties of two-dimensional
silicon carbide}
\author{Arushi Singh}
\email{arushi.phy@iitb.ac.in}

\affiliation{Department of Physics, Indian Institute of Technology Bombay, Powai,
Mumbai 400076, India}
\author{Vikram Mahamiya}
\email{mahamiyavikram@gmail.com}

\affiliation{National Institute for Materials Science (NIMS), 1-1 Namiki, Tsukuba,
Ibaraki 305-0044, Japan}
\affiliation{Department of Physics, Karpagam Academy of Higher Education, Coimbatore
641021, Tamil Nadu India}
\affiliation{Centre for Computational Physics, Karpagam Academy of Higher Education,
Coimbatore 641021, Tamil Nadu, India}
\author{Alok Shukla}
\email{shukla@iitb.ac.in}

\affiliation{Department of Physics, Indian Institute of Technology Bombay, Powai,
Mumbai 400076, India}
\begin{abstract}
Recently, an atomic-scale two-dimensional silicon carbide monolayer
has been synthesized {[}Polley \emph{et al., }Phys. Rev. Lett. \textbf{130},
076203 (2023){]} which opens up new possibilities for developing next-generation
electronic and optoelectronic devices. Our study predicts the pristine
SiC monolayer to have an ``indirect'' band gap of 3.38 eV $(K\rightarrow M)$
and a ``direct'' band gap of 3.43 eV $(K\rightarrow K)$ calculated
using the HSE06 functional. We performed a detailed investigation
of the various possible defects (i.e., vacancies, foreign impurities,
antisites, and their various combinations) on the structural stability,
electronic, and optical properties of the SiC monolayer using a first-principles
based density-functional theory (DFT) and molecular dynamics (MD)
simulations. A number of physical quantities such as the formation
energy, electronic band gap, and the effective masses of charge carriers,
have been calculated. We report that the SiC monolayer has a very
low formation energy of 0.57 eV and can be stabilized on TaC \{111\}
film by performing the surface slab energy and interfacial adhesion
energy calculations. Nitrogen doping is predicted to be the most favorable
defect in silicon carbide monolayer due to its very low formation
energy, indicating high thermodynamic stability. The analysis of the
electronic band structure and the density of states shows that the
additional impurity states are generated within the forbidden region
in the presence of defects, leading to a significant reduction in
the band gap. An interesting transition from semiconducting to metallic
state is observed for $N_{C}$ and $Al_{Si}$ defective systems. For
the pristine SiC monolayer, we find that the conduction band is nearly
flat in the $M\rightarrow K$ direction, leading to a high effective
mass of $3.48m_{o}$. A significant red shift in the absorption edge,
as well as the occurrence of additional absorption peaks due to the
defects, have been observed in the lower energy range of the spectrum.
The calculated absorption spectra span over the visible and ultraviolet
regions in the presence of defects, indicating that the defective
SiC monolayers can have potential optoelectronic applications in the
UV-visible region.
\end{abstract}
\maketitle

\section{Introduction}

\label{sec:introd}

The successful exfoliation of a graphene monolayer from graphite in
2004 \citep{geim2007rise} triggered a significant spike in the research
activity of finding other possible two-dimensional (2D) materials.
Due to their superior physical, chemical, and electronic properties
\citep{PhysRevB.74.075404,BOSTWICK2009380}, 2D materials stand out
as a new category when it comes to the possible commercial application
\citep{hicks1993effect,LIU201999} in the manufacture of flexible
electronic, nanoelectronic, and optoelectronic devices of the next
generation \citep{wang2012electronics,mak2016photonics,zeng2012valley,schedin2007detection}.
Their peculiar properties such as lower scattering rate of charge
carriers, higher carrier mobility and conductivity \citep{PhysRevLett.100.016602,PhysRevLett.97.187401}
make these materials more efficient than bulk silicon in the operation
of electronic devices, which is the basis of the current semiconductor
industry. However, the zero band gap of graphene \citep{novoselov2005two}
is the key issue which limits its potential application in nanoelectronics.
Having a suitable band gap is essential for a 2D material to be a
candidate for device applications. Therefore, many techniques have
been employed to open the band gap of graphene, such as chemical doping,
in-plane strain, functionalization with hydrogen, and molecule adsorption
\citep{boukhvalov2008tuning,cocco2010gap,chang2012band,park2015band,doi:10.1021/nn800459e,sahu2017band,strek2015laser},
etc.. Denis \emph{et al.} \citep{jpcc.5b11709} demonstrated that
the band gap of monolayer and bilayer graphene can be opened with
aluminum, silicon, phosphorus, and sulfur substitutional impurities.
One approach to opening the band gap of graphene is to replace every
second C atom with a Si atom. The resulting SiC monolayer has a stable
planar structure with a band gap of 2.53 eV (using the generalized
gradient approximation -- Perdew-Burke-Ernzerhof functional, GGA-PBE),
as reported in various density functional theory (DFT) based studies
\citep{IOP-alaal2016first,PhysRevB.76.035343}. The electronegativity
difference between Si and C atoms causes polarization of the covalent
bonds in the monolayer giving it a partly ionic character leading
to the opening of the band gap. Also, the dynamical stability of the
2D-SiC monolayer has been confirmed in several previous reports \citep{shi2015predicting,PhysRevB.84.085404,JMCC-fan2017novel}.
A planar SiC monolayer sheet is formed due to the $sp^{2}$ hybridization
between Si and C atoms, as compared to the $sp^{3}$ hybridization
in bulk SiC phase \citep{chabi2021creation,susi2017computational}.
Theoretical studies have also confirmed that the ultrathin wurtzite
structures such as SiC, BeO, GaN, ZnO, and AlN thin films can undergo
phase transformation ($sp^{3}$ to $sp^{2}$) and remain stable. Freeman
\emph{et al. }\citep{freeman2006graphitic} demonstrated that wurtzite
structures such as SiC and ZnO adopt a graphitic structure when thinned
down to very few atomic layers, as it is the most stable structure
for ultrathin materials. Later, this theoretical prediction was experimentally
verified for ultrathin ZnO \citep{tusche2007observation}, AlN \citep{tsipas2013evidence},
and MgO films \citep{goniakowski2004using,goniakowski2007prediction}.
Due to the phase transformation from bulk SiC to a single layer of
SiC, the Si-C bond length is reduced from 1.89 $\text{Å}$ to 1.79
$\text{Å}$, and the bond angle increases from 109$\lyxmathsym{\textdegree}$
to 120$\lyxmathsym{\textdegree}$ \citep{2023vacancyrelated,zhang2020material,susi2017computational},
consistent with its honeycomb structure. Due to the reduced dimensionality
and quantum confinement along one direction, monolayer SiC is expected
to exhibit exotic optical and electronic properties which may find
potential application in the modern-day electronics industry \citep{eddy2009silicon,castelletto2014silicon,chabi2020two}.
As reported in the literature \citep{PhysRevB.80.155453}, the 2D-SiC
has excellent mechanical properties and is found to be one of the
stiffest 2D materials after graphene and h-BN. The SiC monolayer is
highly suitable for its application in high-power and high-temperature
devices due to its relatively wide band gap and superior physical
and thermal properties \citep{choyke2003silicon}.

In terms of experimental efforts, Lin\emph{ }and co-workers \citep{ACS-JPCC.5b04113}
were able to synthesize disordered quasi-2D-SiC flakes (thickness
< 10 nm) through the reactive interaction between silicon quantum
dots and graphene. Although these quasi-2D-SiC flakes were not atomically
thin, the authors did observe changes in the electronic structure
indicative of quantum confinement. Chabi \emph{et al.} were able to
push down the average thickness of 2D-SiC nanoflakes ( average size
$\sim2$ $\mu m$) to 2-3 nm (which estimates between seven to ten
atomic layers) by a catalyst-free carbothermal method \citep{Chabi_2016}.
Very recently, Polley \emph{et al.} \citep{PhysRevLett.130.076203}
reported to have successfully grown an atomically thin large-scale
epitaxial monolayer of honeycomb SiC by employing the ``bottom-up''
solid state fabrication technique. They found that, unlike many other
2D materials that suffer environmental instability, this experimentally
synthesized 2D-SiC monolayer is environmentally stable and does not
degrade at room temperature. 

It is well known that the defects (dopants, vacancies, antisites,
interstitials, etc.) are unavoidable during fabrication or post-processing
of materials. These defects play a crucial role in significantly tailoring
the electronic and optical characteristics of the 2D material. Due
to the quantum confinement effect of both host and defected wave functions
in 2D materials, the impact of defects on the electronic and optical
properties may considerably differ from their three-dimensional (3D)
bulk counterpart. In this paper we undertake a systematic study of
structural stability, electronic, and optical properties of some favorable
defects in 2D-SiC monolayer utilizing first-principles-based DFT and
MD simulations. We have investigated the point defects (i.e., vacancy,
antisites), foreign impurity defects (i.e., boron, phosphorus, nitrogen
and aluminum doping), and complex defects (i.e., doping-vacancy, antisite-vacancy),
which are a combination of the aforementioned defects. 

The remainder of this paper is organize as follows. In the next section
we briefly describe our computational approach, followed by the presentation
and discussion of our results in Sec. \ref{sec:Results-and-analysis}.
Finally, in Sec. \ref{sec:conclusion} we present our conclusions.

\section{Computational methods\label{sec:Method}}

All the calculations presented in this work have been performed within
the framework of the first-principles DFT as implemented in the Vienna
\emph{Ab initio} Simulation Package (VASP) code \citep{kresse1996efficiency,kresse1999ultrasoft}.
In this approach the electron-ion interactions are described using
the projected augmented wave (PAW) \citep{PhysRevB.50.17953} method
with the energy cutoff set at 500 eV. The screened hybrid functional
of Heyd--Scuseria--Ernzerhof (HSE06) \citep{heyd1242006,krukau2006influence}
with $25\%$ of the Hartree--Fock exchange and $0.2\text{ Å}^{-1}$
screening parameter has been employed for accurate band structure
calculations. The Brillouin zone is sampled using the Monkhorst-Pack
scheme, with an automatically generated k-point mesh of $5\times5\times1$
and $7\times7\times1$ for relaxation and single-point energy calculations,
respectively. The convergence threshold for the forces felt by atoms
is chosen to be 10$^{-2}$ eV/$\lyxmathsym{\AA}$, and for the total
energy minimization, it is 10$^{-5}$eV. For performing calculations
on defective configurations, a $4\times4$ supercell of SiC monolayer
is chosen. A vacuum space of 15 $\lyxmathsym{\AA}$ is kept along
the z-axis to eliminate the interactions between the periodic images
of the supercell. The surface matching for the SiC monolayer and TaC
\{111\} slab is performed using the Zur and McGill algorithm \citep{zur1984lattice},
which builds the lowest area supercell that matches the surface area
of two lattices. In the slab and interface calculations, both lattice
vectors and atomic positions are optimized. The van der Waals dispersion
corrections are employed by using Grimme\textquoteright s DFT-D3 method
\citep{grimme2010consistent}. We have also considered dipole corrections
in the calculation of the total energies of the surface slabs. The
structural integrity at room temperature for the pristine as well
as the defect-induced monolayer is verified by performing \emph{ab-initio}
molecular dynamics simulations (AIMD) \citep{doi:10.1080/00268978400101201}.
The AIMD simulations are performed in two consecutive steps: First,
the system is kept in a micro-canonical ensemble (NVE) for 6 ps of
time duration using time steps of 1 fs each, while the temperature
is raised from 0 to 300 K. Next, we kept the system in a canonical
ensemble (NVT) at 300 K for 6 ps of time duration by employing a Nose--Hoover
thermostat. For optical calculations, the frequency-dependent imaginary
part of the dielectric function is obtained from the momentum matrix
elements between the occupied valence bands and the unoccupied conduction
bands, while the real part follows from Kramers-Kronig relations \citep{kronig1926theory}.

\section{Results and analysis \label{sec:Results-and-analysis}}

\begin{figure}
\includegraphics[scale=0.3]{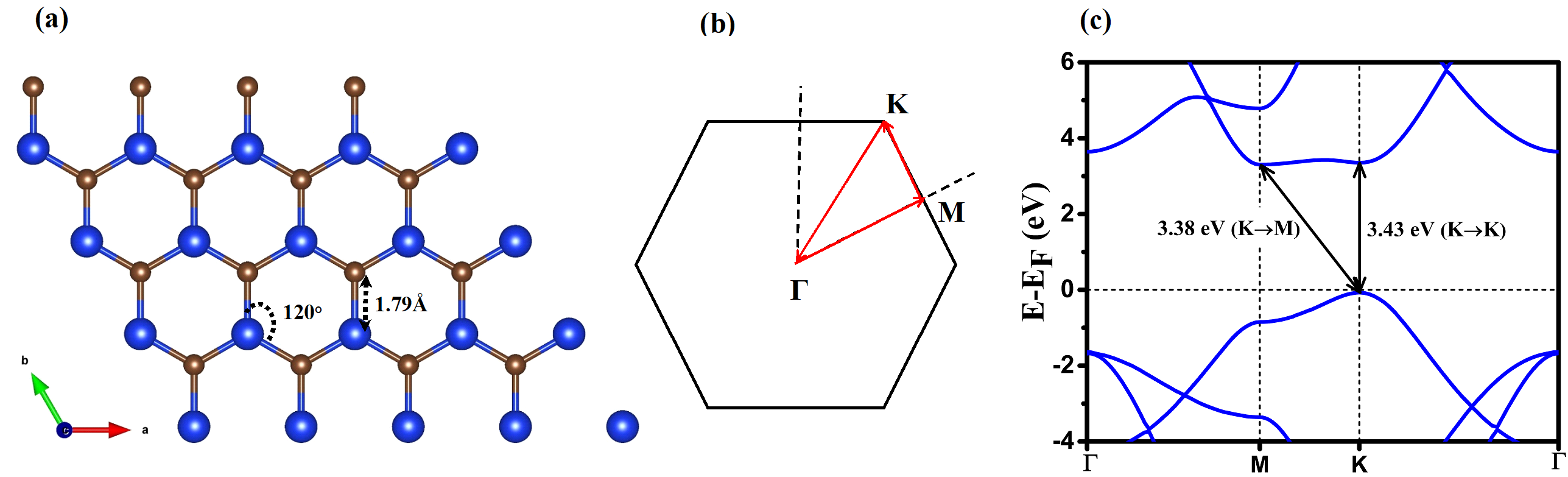}

\caption{\textcolor{red}{\label{fig:fig-1}}(a) Optimized structure of 4 \texttimes{}
4 supercell of SiC monolayer with Si-C bond length of 1.79 $\text{Å}$,
and Si-C-Si (and C-Si-C) bond angles 120$\lyxmathsym{\protect\textdegree}${[}the
blue (brown) spheres denote the Si (C) atoms, and the monolayer is
assumed to lie in the $xy$ plane{]}. (b) $\Gamma$-M-K-$\Gamma$
path in the corresponding reciprocal space and the (c) calculated
HSE06 electronic band structure of a primitive cell of 2D-SiC monolayer.}
\end{figure}

The valence shells of C and Si have electronic configurations of $(2s^{2},2p^{2})$
and $(3s^{2},3p^{2})$, respectively. Thus, in the planar structure
of the SiC monolayer, both the atoms participate in sp$^{2}$--p$_{z}$
hybridization, so that the nearest-neighbor atoms (C and Si) are bonded
together by $\sigma$ bonds (Fig \ref{fig:fig-1}), while the p$_{z}$
orbitals of each atom give rise to the $\pi$/$\pi^{*}$ bands corresponding
to the valence/conduction bands of the system. Our calculations started
by relaxing the pristine SiC monolayer, leading to the optimized bond
length of 1.79 $\lyxmathsym{\AA}$ and optimized Si-C-Si (and C-Si-C)
bond angle equal to 120° (Fig.\ref{fig:fig-1}(a)), fully consistent
with the previous theoretical studies \citep{PhysRevB.80.155453,2023vacancyrelated,JPCC,susi2017computational}.
The nature of the band gap in SiC monolayer is debatable in the reported
literature so far. The ``direct'' band gap has been predicted by
Lin \emph{et al. }\citep{lin2013ab} and Ferdous \emph{et al.} \citep{ferdous2019tunable}
using SIESTA \citep{soler2002siesta} and Quantum Espresso \citep{giannozzi2017advanced,giannozzi2009quantum}
simulation codes, respectively, employing the GGA-PBE exchange-correlation
functional. Guo \emph{et al.} \citep{PhysRevB.76.035343} and Hou
\emph{et al.} \citep{cite-60} have also predicted a ``direct''
band gap for the SiC monolayer using the LDA and GGA-PW91 functional,
respectively, as implemented in VASP. However, Hsueh and co-workers
\citep{PhysRevB.84.085404} reported that the band gap is ``indirect''
using the local density approximation (LDA) functional and transforms
to ``direct'' upon employing GW self-energy corrections. Zheng \emph{et
al.} \citep{lu2012tuning} have also claimed an ``indirect'' band
gap using LDA and GWA functionals as implemented in the ABINIT code\citep{gonze2002first}.
Ciraci \emph{et al.} \citep{PhysRevB.80.155453} also reported an
\textquotedblleft indirect\textquotedblright{} band gap based on their
LDA calculations. It is commonly known that LDA/GGA functionals underestimate
the band gaps in most materials. Therefore, to obtain a more accurate
estimate of the band gap of the SiC monolayer, we decided to perform
calculations using the hybrid functional HSE06. For this purpose,
we used the geometry optimized using the PBE functional, and performed
HSE06-based band structure calculations on the primitive cell of SiC
monolayer containing two atoms. We found an ``indirect'' band gap
of 3.38 eV $(K\rightarrow M)$ and a ``direct'' band gap of 3.43
eV $(K\rightarrow K)$ in the SiC monolayer, as demonstrated in Fig.
\ref{fig:fig-1}(c). For the sake of comparison, we have also computed
the band structure using the PBE functional (see Fig. S1 \citep{SupplementalMaterial}),
and found that the system exhibits an indirect band gap of 2.55 eV
$(K\rightarrow M)$ and a direct one of 2.56 eV $(K\rightarrow K)$.
Upon comparing the two band structures, we conclude that they differ
quantitatively with the HSE06 band gaps, being $\approx$ 0.8 eV larger;
however, their qualitative features are very similar.

\begin{figure}
\includegraphics[scale=0.3]{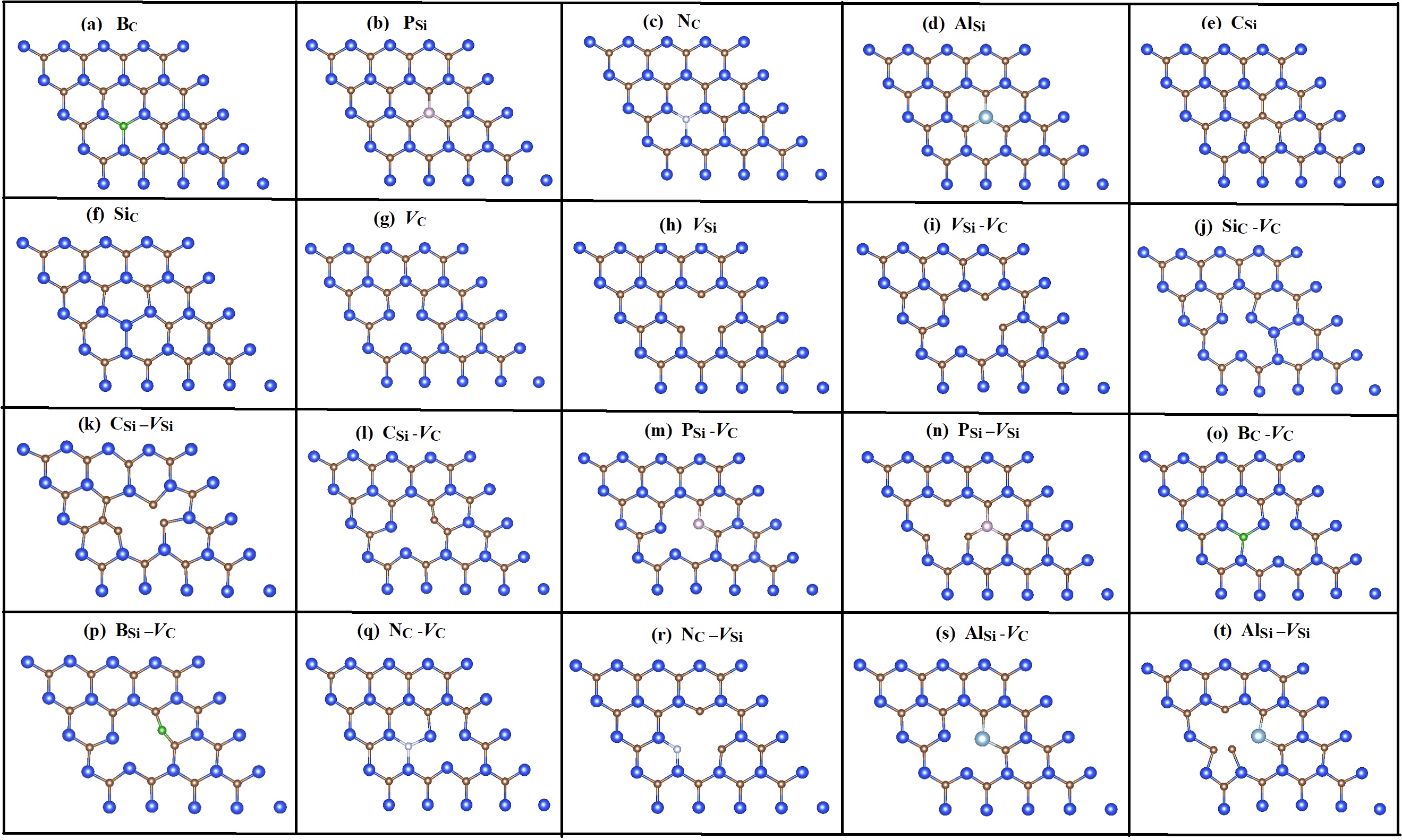}

\caption{Relaxed ground state geometry of point defects of monolayer SiC considered
in this work. Simple defects: (a) $B_{C}$, (b) $P_{Si}$, (c) $N_{C}$,
(d) $Al_{Si}$, (e) $C_{Si}$, (f) $Si_{C}$, (g) $V_{C}$, and (h)
$V_{Si}$; Complex defects: (i) $V_{Si}$-$V_{C}$, (j) $Si_{C}$-$V_{C}$,
(k) $C_{Si}$-$V_{Si}$, (l) $C_{Si}$-$V_{C}$, (m) $P_{Si}$-$V_{C}$,
(n) $P_{Si}$-$V_{Si}$, (o) $B_{C}$-$V_{C}$, (p) $B_{Si}$-$V_{C}$,
(q) $N_{C}$-$V_{C}$, (r) $N_{C}$-$V_{Si}$, (s) $Al_{Si}$-$V_{C}$,
and (t) $Al_{Si}$-$V_{Si}$ (all the defective monolayers are assumed
to lie in the $xy$ plane). \label{fig: fig-2}}
\end{figure}

Due to the slight difference in the relative energies in the conduction
band edge at the K and M points of the Brillouin zone, the nature
of the band gap and the related properties can be easily tuned with
the introduction of various types of defects, impurities, strains,
etc.. Hassanzada \emph{et al.} \citep{PhysRevB.102.134103} proposed
that the Stone-Wales-related defects could be promising hosts for
single-photon quantum emitters. The paramagnetic color centers are
connected to silicon vacancy-related defects, as reported by Mohseni
and co-workers \citep{2023vacancyrelated}. The effects of various
substitutional impurities, adatoms, and vacancy defects on magnetic
properties have been investigated by Ciraci \emph{et al.} \citep{PhysRevB.81.075433}.
Despite a fair number of studies on 2D-SiC monolayer, a comprehensive
study of the stability of various defective structures and their electronic
and optical characteristics is still lacking. Therefore, we present
here a detailed investigation of the structural stability and optoelectronic
properties of SiC monolayer with following types of defects: (i) substitutional
defects, where a dopant X can replace one Si atom or a dopant Y can
replace one C atom, i.e. $X_{Si}$ or $Y_{C}$, (ii) vacancy, where
either a Si or C atom or both the atoms are missing from their respective
sites i.e. $V_{Si}$ or\emph{ }$V_{C}$ or $V_{Si}$-$V_{C}$, (iii)
antisites, where a Si atom is present at the C atom site or vice-versa
i.e. $Si_{C}$ or $C_{Si}$, (iv) antisite-vacancy, a combination
of both (ii) and (iii), where either a C or Si atom is missing from
their respective sites forming a vacancy defect, while at the same
time, C atom is present at the Si-site or vice-versa, i.e., $Si_{C}$-$V_{Si/C}$
or $C_{Si}$-$V_{Si/C}$; and (v) doping-vacancy, a combination of
both (i) and (ii), i.e., $X_{Si}$-$V_{Si/C}$ or $Y_{C}$-$V_{Si/C}$.
The ground state geometries of these defects in a 4$\times$4 supercell
of SiC monolayer are presented in Fig \ref{fig: fig-2}. 

We have divided the remainder of this section into four parts to discuss
the effects of defects on (a) structural stability, (b) electronic
structure, (c) effective mass of charge carriers, and (d) optical
properties. 

\subsection{Stability of defects in 2D-SiC monolayer: \textit{Ab Initio} Atomistic
Thermodynamics and Molecular Dynamics calculations}

In this section, we first probe the thermodynamics of various defect-induced
SiC monolayer systems by calculating their formation energies in both
Si-rich and C-rich environments, which is crucial in supporting the
experimental work to improve the setup conditions during the growth
of the 2D-SiC monolayer. The structural stability of the defective
systems is probed by performing AIMD simulations at room temperature. 

\begin{figure}
\includegraphics[scale=0.5]{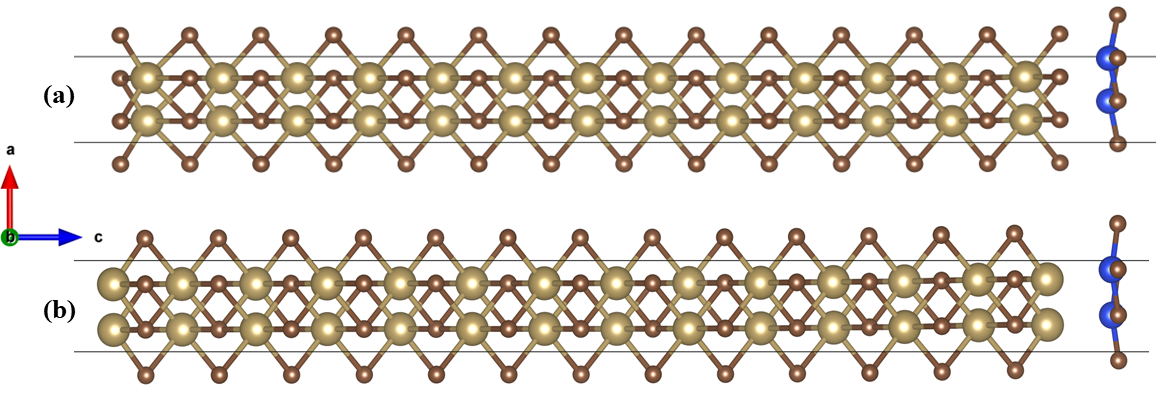}\caption{Interface model optimized structures of 3.3 nm thick TaC \{111\}-SiC
monolayer for (a) C-terminated TaC \{111\} slab and (b) Ta-terminated
TaC \{111\} slab\label{fig:fig-3} }

\end{figure}

The formation energy of monolayer SiC is obtained using the formula
$\Delta H_{F}$ = $\mu_{SiC}-(\mu_{Si}+\mu_{C})$, where $\mu_{SiC}$
is the chemical potential of the monolayer SiC unit cell, while $\mu_{Si}(\mu_{C})$
denote the chemical potential of the stable phase of Si (C). The calculated
formation energy of the monolayer is found to be 0.57 eV, which is
in excellent agreement with the values reported in the literature
\citep{PhysRevB.102.134103}. The low positive value of formation
energy indicates that the SiC monolayer is a metastable phase of silicon
carbide and hence could be stabilized experimentally. To investigate
the role of substrate in the growth of SiC monolayer, we have performed
substrate-film adhesion energy calculations using the first-principles
DFT, combined with the substrate interface theory. Experimentally,
an epitaxial monolayer of 2D SiC was recently synthesized  on ultrathin
rocksalt cubic TaC\{111\} surface of thickness $\sim$3nm, which was
accommodated on a 4H-SiC substrate \citep{PhysRevLett.130.076203}.
We have extracted a 3.3 nm thick \{111\} Miller surface of rocksalt
cubic TaC from the bulk TaC cubic halite, and performed the full geometry
optimization and self-consistent field calculations. The surface slab
energies of TaC\{111\} slab and SiC monolayer are calculated using
the Eq. (\ref{eq:surface slab energy}) \citep{cite-67,cite-68}.

\begin{equation}
\sigma=\frac{1}{2A}[E_{slab}-(NE_{Bulk}+n\mu_{i})],\label{eq:surface slab energy}
\end{equation}

where $E_{slab}$ and $E_{Bulk}$ are the total energies of the surface
slab and bulk units, $N$, $n$, $A$, and $\mu_{i}$ are, respectively,
the number of formula units, number of extra atoms in surface slab,
surface area, and chemical potential of elementary atoms. $\mu_{C}$,
$\mu_{Ta}$, and $\mu_{Si}$ are calculated from their stable bulk
phases under both Ta-rich and C-rich conditions, since the monolayer
SiC is grown on the TaC film. The surface energy of monolayer SiC
is found to be 11.55 $J/m^{2}$, computed with respect to the stable
4H phase of SiC. The surface energies of the 3.3 nm thick TaC \{111\}
slab $(\sigma_{TaC\{111\}})$ with C and Ta-terminations are provided
in Table 1. We found that Ta-terminated TaC\{111\} slabs have lower
surface energy compared to C-terminated TaC\{111\}, which is consistent
with the previous slab energy calculations of various rocksalt transition-metal
carbide facets by Quesne \emph{et al.}\citep{quesne2018bulk}. Next,
we construct two interfaces of TaC \{111\} slab and SiC monolayer
corresponding to the C and Ta terminations of TaC. The silicon carbide
monolayer is kept at a distance 2 Å above the metal carbide\textquoteright s
surface, and the optimized distance between the C(Ta) terminated TaC
\{111\} and Si atom of monolayer SiC is 2.36 Å (2.72 Å). The optimized
structures of the TaC \{111\}-SiC interface, with the C and Ta terminations
are presented in Figs. \ref{fig:fig-3}(a) and \ref{fig:fig-3}(b).
We report a small lattice mismatch of 1.87 \% in the interface structures
composed of TaC \{111\} 3.3 nm thick film and SiC monolayer. Next,
we calculate the bonding strengths of the atoms present at the TaC
\{111\}-SiC interface and the interface stability by calculating the
interfacial adhesion energy $(E_{adh})$ using Eq. (\ref{eq:interfacial adhesion energy})\citep{restuccia2023high}

\begin{equation}
E_{adh}=\frac{1}{A}[E_{TaC\{111\}-SiC}-E_{SiC}-E_{TaC\{111\}}].\label{eq:interfacial adhesion energy}
\end{equation}

Here, $E_{SiC}$, $E_{TaC\{111\}}$, and $E_{TaC\{111\}-SiC}$ are
the total energies of the SiC monolayer, TaC\{111\} slabs, and TaC\{111\}-SiC
interfaces, respectively, and $A$ is the lateral interface area.
The adhesion energy for the TaC \{111\}-SiC monolayer was found to
be -1.44 $J/m^{2}$ and -2.05 $J/m^{2}$ for C and Ta-terminated TaC
\{111\} facets. The negative values of adhesion energies imply that
TaC \{111\} substrate can stabilize the SiC monolayer. The adhesive
energy of the Ta-terminated TaC \{111\}-SiC interface is less than
that of the C-terminated TaC \{111\}-SiC interface, because Ta-terminated
TaC \{111\} facet is more stable than the C-terminated TaC \{111\}.

\begin{table}
\begin{tabular}{|c|c|c|c|c|c|}
\hline 
\multicolumn{4}{|c|}{$\sigma_{TaC\{111\}}(J/m^{2})$ } & \multicolumn{2}{c|}{$E_{adh}(J/m^{2})$}\tabularnewline
\hline 
\multicolumn{2}{|c|}{Ta-termination} & \multicolumn{2}{c|}{C-termination} & \multirow{2}{*}{Ta-termination} & \multirow{2}{*}{C-termination}\tabularnewline
\cline{1-4} \cline{2-4} \cline{3-4} \cline{4-4} 
C-rich & Ta-rich & C-rich & Ta-rich &  & \tabularnewline
\hline 
3.10 & 1.98 & 5.26 & 6.34 & -2.05 & -1.44\tabularnewline
\hline 
\end{tabular}

\caption{Surface $(\sigma)$ and interfacial adhesion $(E_{adh})$ energy of
TaC \{111\} facet and SiC monolayer with different termination and
chemical environments. \label{tab:table-1}}

\end{table}

The cohesive energy of an isolated pristine SiC monolayer is calculated
as $E_{C}=E_{T}^{SiC}-(E_{T}^{Si}+E_{T}^{C})$ in terms of the optimized
total energy of the unit cell of SiC monolayer and total energies
of isolated Si and C atoms, respectively. The calculated cohesive
energy is found to be -11.87 eV, indicating the high stability of
the material.

We investigated the relative stability of various possible defects
under investigation by computing their formation energies. The formation
energy of a neutral defect, $E_{F}(D)$ is calculated using Eq. (\ref{eq:formation-energy-1})
\citep{PhysRevLett.67.2339,RevModPhys.86.253}. 
\begin{equation}
E_{F}(D)=E_{total}(D)-E_{total}(host)+\sum_{i}n_{i}\mu_{i}\label{eq:formation-energy-1}
\end{equation}

\begin{figure}
\includegraphics[scale=0.65]{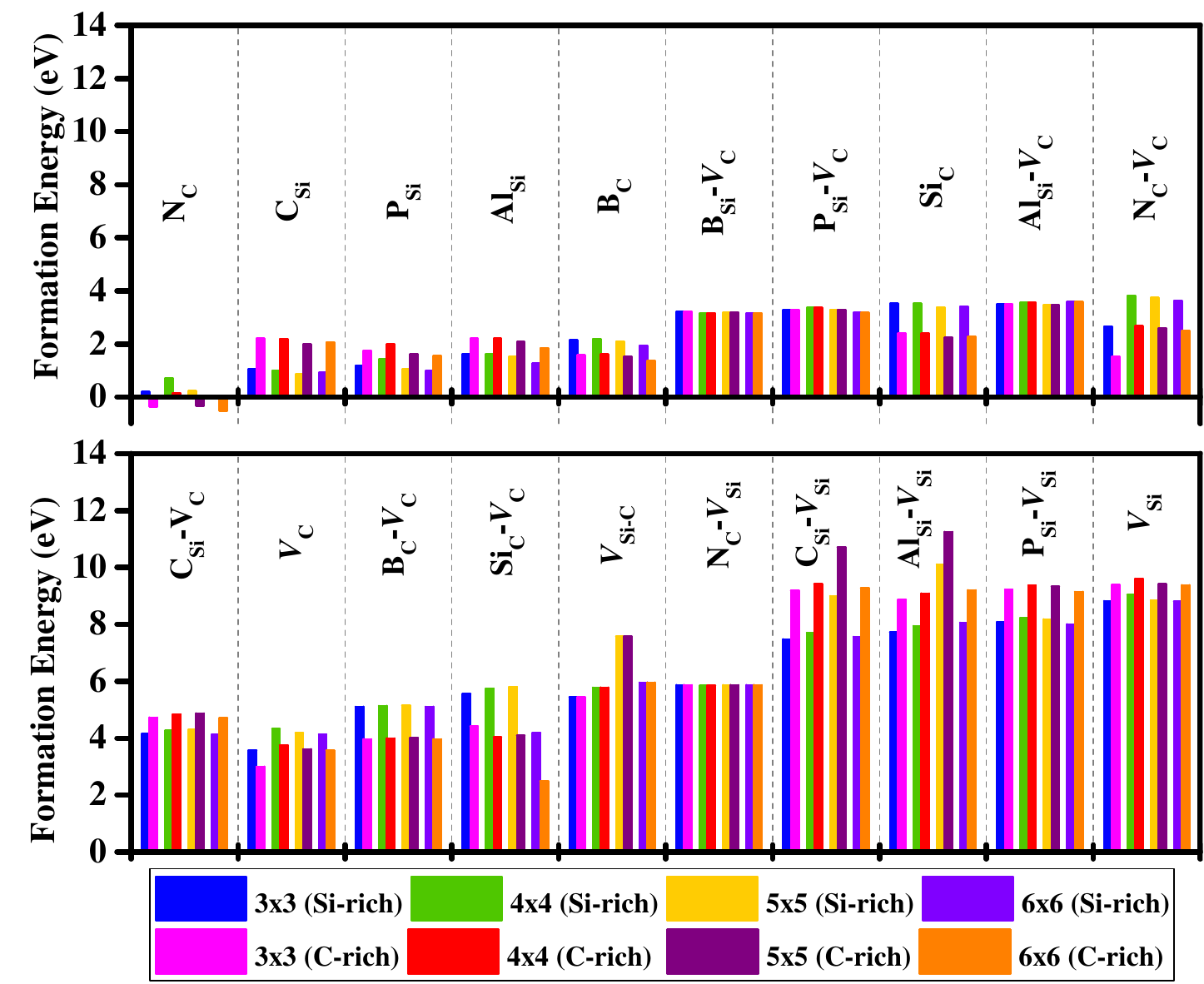}

\caption{\label{fig:fig-4}Defect formation energies (in eV) calculated for
(3\texttimes 3), (4\texttimes 4), (5\texttimes 5) and (6\texttimes 6)
supercell in Si-rich (C-poor) and C-rich (Si-poor) growth conditions
for various possible neutral defects in the 2D-SiC monolayer.}

\end{figure}

Above $E_{total}(D)$ and $E_{total}(host)$ represent the total energies
of the optimized structures of the defective, and pristine monolayers,
respectively. $n_{i}$ is the number of atoms removed from ($n_{i}>0)$
or added to ($n_{i}<0$) the SiC monolayer to form a defect, and $\mu_{i}$
is their corresponding chemical potential. The chemical potential,
$\mu_{i}$ depends on the chemical environment of the system, therefore
we have calculated the formation energies of defects in both Si-rich
(C-poor) and C-rich (Si-poor) growth conditions. $\mu_{i}$ of the
constituent atoms are calculated from their stable bulk phases. The
calculated formation energies of various possible defective systems
considered in this work are shown in Fig. \ref{fig:fig-4} for both
Si-rich (C-poor) and C-rich (Si-poor) growth conditions.

Here, the formation energy is calculated using the finite-size supercell
methodology, where a defect is introduced in a suitably large simulation
supercell to minimize the defect-defect interaction, which can have
a significant impact on the defect formation energy. For charged defects,
the interactions between the periodic images of defects are significant
due to the presence of long-range electrostatic potential term that
converges very slowly as a function of supercell size. However, for
neutral defects, which are explored in this work, the defect-defect
interactions are negligible. Therefore the formation energy converges
fairly quickly with the increasing supercell size. As a result, the
simulations employing smaller supercells can yield fairly precise
estimates for the formation energy of the neutral defects. In this
work, we have computed the defect formation energy using a (4 \texttimes{}
4) supercell and also compared it with the formation energy obtained
using different supercell sizes, i.e., (3 \texttimes{} 3), (5 \texttimes{}
5) and (6 \texttimes{} 6) (Fig. \ref{fig:fig-4}). We found that the
trend in the formation energy of various defects considered in the
present study remains unaffected by the different supercell sizes
(Fig. \ref{fig:fig-4}). Additionally, the formation energy is very
well converged for the majority of the defects which makes the (4
\texttimes{} 4) supercell size a suitable choice to further study
the defect properties.

The calculated theoretical formation energy of the $N_{C}$ defect
is the lowest for the C-rich growth condition, indicating that this
defect is most likely to be found in the SiC monolayer. The defect
formation energy of $V_{C}$ is lower than that of $V_{Si}$, which
is similar to the case corresponding to the bulk 4H-SiC phase \citep{4H-SiC}.
The formation energies of $N_{C}$-$V_{Si}$, $Al_{Si}$-$V_{Si}$,
$P_{Si}$-$V_{Si}$, and $C_{Si}$-$V_{Si}$ are greater than those
of $N_{C}$-$V_{C}$, $Al_{Si}$-$V_{C}$, $P_{Si}$-$V_{C}$, and
$B_{Si}$-$V_{C}$ . The large formation energies associated with
the Si vacancy are expected, because the Si-atom is heavier than the
C-atom, and hence will be difficult to remove. Another point that
can be noted is that the formation energy of the $C_{Si}$ type defect
is lower than the $Si_{C}$ type defect because C atoms prefer the
sp$^{2}$ configuration more than the Si atom does. Among the dopant-type
defects, $Al_{Si}$ and $B_{C}$ have greater formation energies than
$N_{C}$ and $P_{Si}$, which can be attributed to the atomic size
and electronegativity of the dopant atoms. To further reveal the thermal
stability of pristine and defective systems at room temperature (300
K), we performed \emph{ab-initio} molecular dynamics (AIMD) simulations.
The maximum bond length variation is found to be around $\sim4.5\%$
for a pristine SiC monolayer, which is not significant and corresponds
to only $\text{0.08 Å}$. It has been observed that for the doped,
antisite, vacancy and complex defect cases, the maximum bond length
fluctuations are $\sim4.4\%,5.0\%,6.6\%,$ and $6.1\%$ for $N_{C}$,
$Si_{C}$, $V_{Si}$, and $N_{C}$-$V_{Si}$ defects, respectively.
Thus, we conclude that both the pristine as well as defective SiC
monolayers are stable at room temperature.

\subsection{The influence of defects on the electronic structure and magnetic
properties}

\begin{figure}
\includegraphics[scale=0.6]{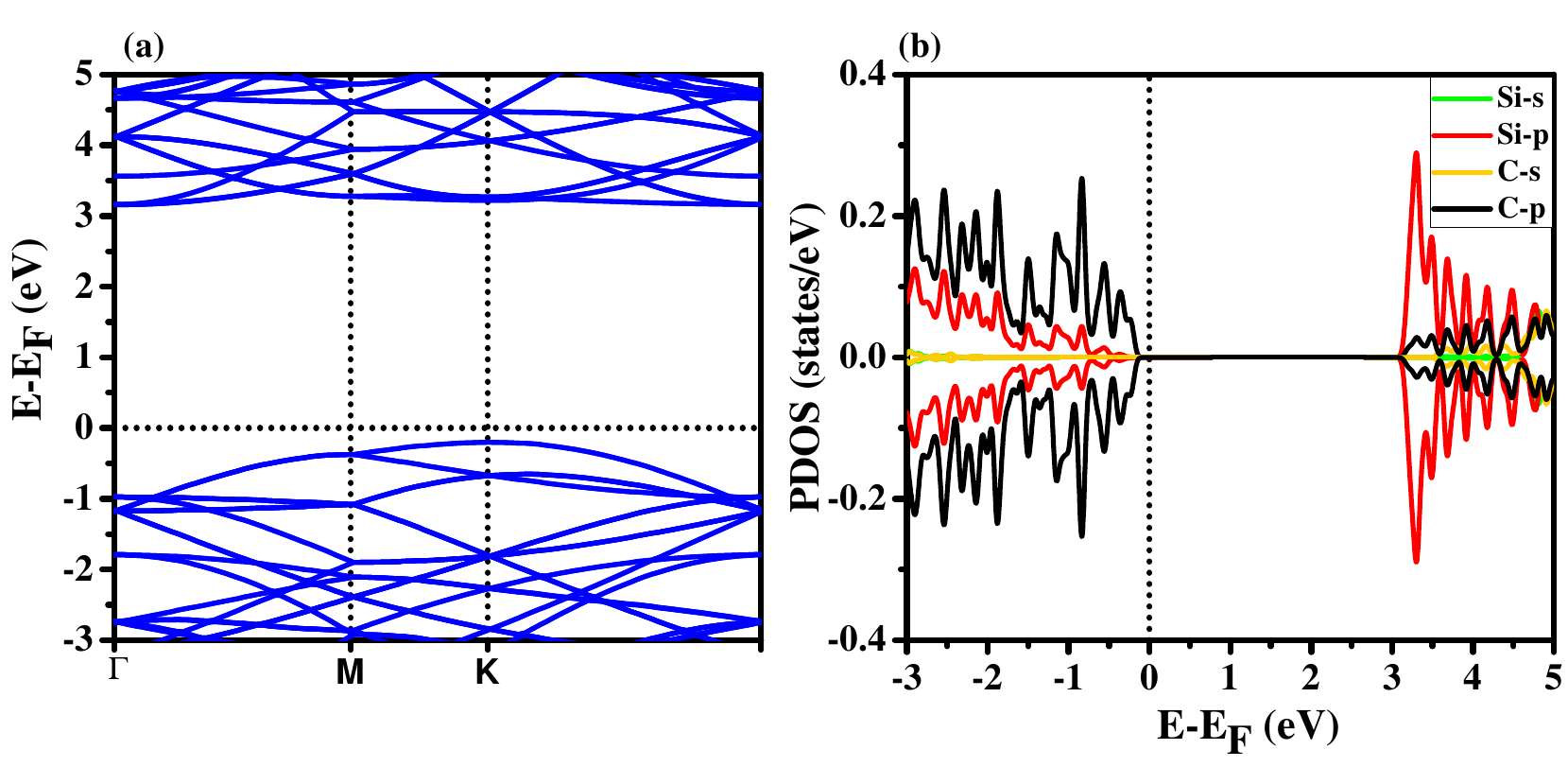}

\caption{\label{fig:fig-5}The calculated (a) band structure, and (b) atom-projected
PDOS of the 4 \texttimes{} 4 supercell of the SiC monolayer. The Fermi
level has been set to zero.}
\end{figure}

This section is devoted to the analysis of the electronic band structure
and electronic density of states (DOS) of various structures under
investigation. As a benchmark test for comparison, we first investigated
the electronic band structure of a 4 \texttimes{} 4 supercell of the
pristine SiC monolayer. Notably, more bands populate the Brillouin
zone because of the band folding caused by a larger number of atoms
in the supercell as compared to the primitive cell. We also examined
the spin-polarized atom-projected electronic density of states (PDOS)
to explore the contribution of each atom and orbital to the valence
and conduction bands in the proximity of the Fermi level. The p-orbital
of the C-atom clearly dominates the valence band, whereas the Si atom's
p-orbital mostly contributes to the conduction band as seen in Fig.
\ref{fig:fig-5}(b). 

\begin{figure}
\includegraphics[scale=0.6]{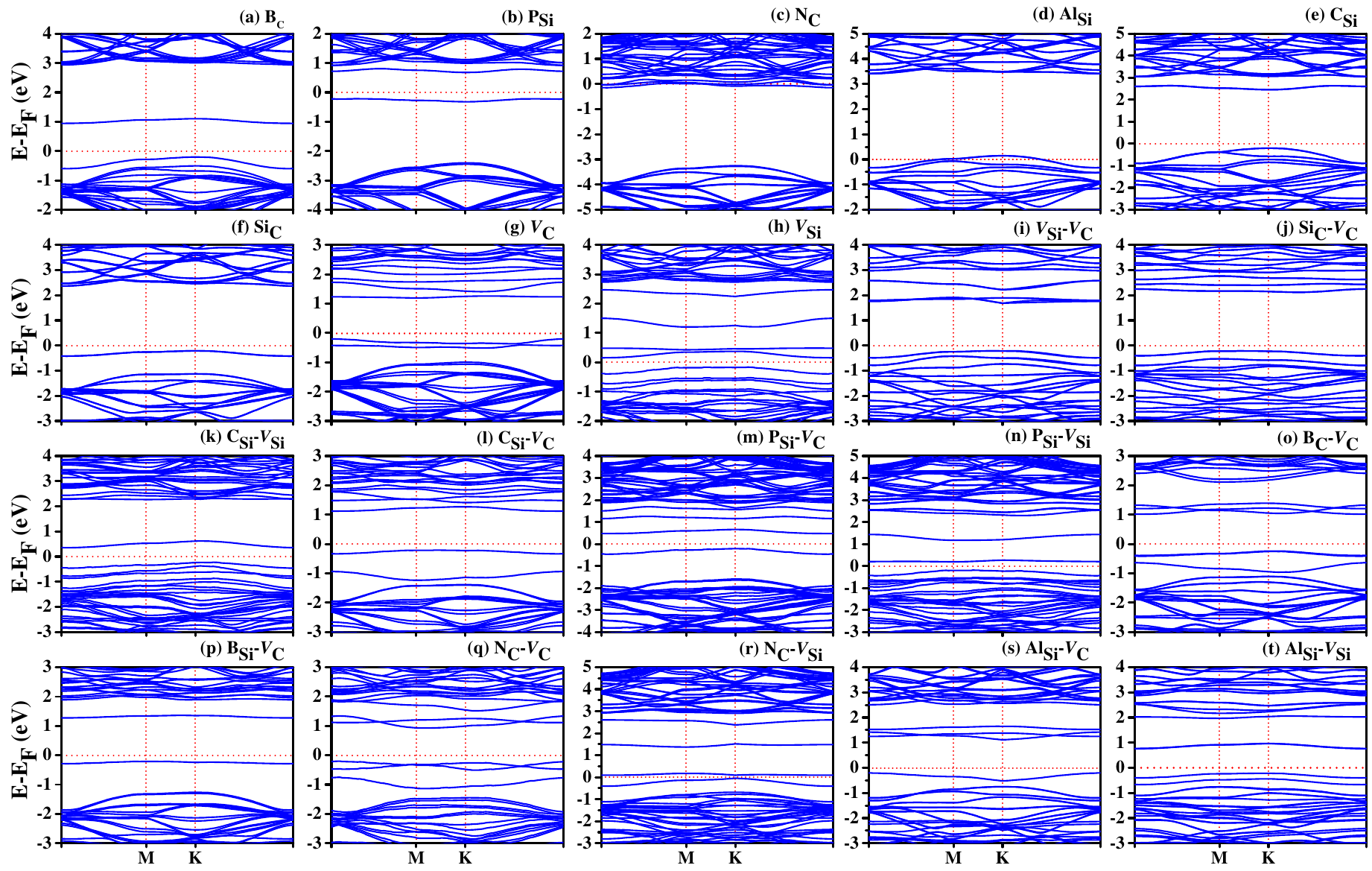}

\caption{\label{fig:fig-6}Electronic band structures calculated using HSE06
for various defects under investigation. Simple defects: (a) $B_{C}$,
(b) $P_{Si}$, (c) $N_{C}$, (d) $Al_{Si}$, (e) $C_{Si}$, (f) $Si_{C}$,
(g) $V_{C}$, and (h) $V_{Si}$. Complex defects: (i) $V_{Si}$-$V_{C}$,
(j) $Si_{C}$-$V_{C}$, (k) $C_{Si}$-$V_{Si}$, (l) $C_{Si}$-$V_{C}$,
(m) $P_{Si}$-$V_{C}$, (n) $P_{Si}$-$V_{Si}$, (o) $B_{C}$-$V_{C}$,
(p) $B_{Si}$-$V_{C}$, (q) $N_{C}$-$V_{C}$, (r) $N_{C}$-$V_{Si}$,
(s) $Al_{Si}$-$V_{C}$, and (t) $Al_{Si}$-$V_{Si}$.}

\end{figure}

\begin{figure}
\includegraphics[scale=0.6]{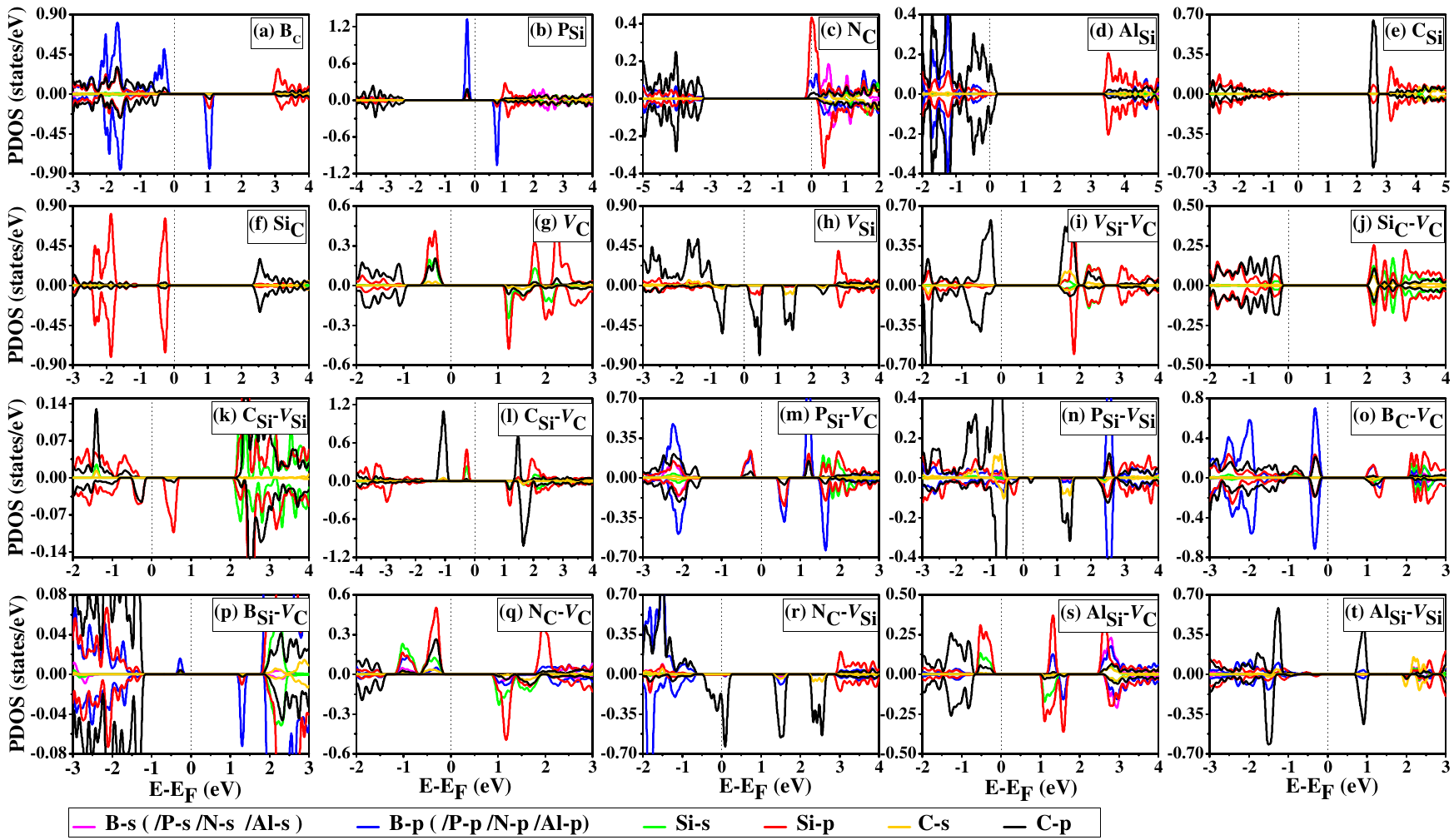}

\caption{\label{fig:fig-7}Atom projected density of states (PDOS) calculated
using HSE06 for various defects under investigation. Simple defects:
(a) B$_{C}$, (b) P$_{Si}$, (c) N$_{C}$, (d) Al$_{Si}$, (e) C$_{Si}$,
(f) Si$_{C}$, (g) V$_{C}$, and (h) V$_{Si}$; Complex defects: (i)
\emph{V}$_{Si}$-\emph{V}$_{C}$, (j) Si$_{C}$-\emph{V}$_{C}$, (k)
C$_{Si}$-\emph{V}$_{Si}$, (l) C$_{Si}$-\emph{V}$_{C}$, (m) P$_{Si}$-\emph{V}$_{C}$,
(n) P$_{Si}$-\emph{V}$_{Si}$, (o) B$_{C}$-V$_{C}$, (p) B$_{Si}$-\emph{V}$_{C}$,
(q) N$_{C}$-\emph{V}$_{C}$, (r) N$_{C}$-\emph{V}$_{Si}$, (s) Al$_{Si}$-\emph{V}$_{C}$,
and (t) Al$_{Si}$-\emph{V}$_{Si}$.}
\end{figure}

To elucidate the role of defects on the electronic structure of defected
2D-SiC, we have performed band structure and PDOS calculations using
hybrid HSE06 functional (refer to Figs. \ref{fig:fig-6} and \ref{fig:fig-7}).
It is interesting to note that the additional defect states are generated
near the Fermi level for $B_{C}$ and $P_{Si}$ type defects as compared
to the pristine monolayer {[}see Figs. \ref{fig:fig-6}(a) and \ref{fig:fig-6}(b){]}.
This leads to a significant reduction in the band gap $(E_{g})$,
which becomes 1.15 eV for $B_{C}$ and 0.90 eV for $P_{Si}$, respectively.
The strong peak of the p-orbital of the dopant atom, i.e., B (P),
is observed near the Fermi level for $B_{C}$ $(P_{Si})$ defected
systems as shown in Figs. \ref{fig:fig-7}(a) and \ref{fig:fig-7}(b),
which confirms that these additional states in the band structure
are arising due to the dopant atom. 

However, for $N_{C}$ $(Al_{Si})$ defects, as shown in Figs. \ref{fig:fig-6}(c)--
\ref{fig:fig-6}(d) (band structure) and Figs. \ref{fig:fig-7}(c)--\ref{fig:fig-7}(d)
(PDOS), the Fermi level lies near the bottom (top) of the conduction
(valence) band within the continuous density of states and hence shows
the interesting defect-induced transition from the semiconducting
to metallic state. The metallicity is basically due to the shift of
the Fermi level of the supercell caused by an extra valence electron
in case of $N_{C}$ and one less electron in case $Al_{Si}$ as compared
to the pristine supercell. From Figs. \ref{fig:fig-7}(c) and \ref{fig:fig-7}(d)
it is obvious that the dominant contribution to the PDOS near the
Fermi level is due to the p orbital of the Si (C) atom for $N_{C}$
$(Al_{Si})$ defects. 

For the antisite defects, i.e., $C_{Si}$ and $Si_{C}$ defects, only
one defect state corresponding to the p orbital of C and Si, respectively,
is found in the midgap region {[}see Figs. \ref{fig:fig-6}(e), \ref{fig:fig-6}(f)
and \ref{fig:fig-7}(e), \ref{fig:fig-7}(f){]}. Therefore, the system
is expected to behave as n-type (p-type) semiconductor under the presence
of $C_{Si}$ ($Si_{C}$) defects with the reduced band gap $(E_{g})$
of 2.66 eV (2.58 eV). 

In the case of a single C (Si)-atom vacancy, i.e., $V_{C}$ $(V_{Si})$,
the atoms nearest to the vacancy defect have an unpaired electron
that forms an unsaturated covalent bond and causes the impurity bands
to emerge in the forbidden region exhibiting a band gap $(E_{g})$
of 1.40 eV (0.34 eV) {[}see Figs. \ref{fig:fig-6}(g) and \ref{fig:fig-6}(h){]}.
A detailed examination from PDOS {[}Figs. \ref{fig:fig-7}(g) and
\ref{fig:fig-7}(h){]} clearly demonstrates that the p orbital of
Si (C) atoms for $V_{C}$ $(V_{Si})$ type defect are the primary
cause of the impurity states. The result contrasts with the bulk 4H-SiC
phase \citep{4H-SiC}, where impurity states created due to a single
Si or C atom vacancy cross the Fermi level, inducing the metallicity
in the system. The mid-gap defect states are also observed in the
electronic structure of the $V_{Si}$-$V_{C}$ double vacancy, mainly
due to the p orbital of the C atom, as is obvious from Figs. \ref{fig:fig-6}(i)
and \ref{fig:fig-7}(i). This leads to the reduced band gap $(E_{g})$
of 1.90 eV.

The symmetric character of PDOS suggests the non-magnetic nature of
the pristine SiC monolayer {[}Fig. \ref{fig:fig-5}(b){]}. However,
some defects can induce a net magnetic moment due to imbalanced up
and down spin channels, as observed in Fig. \ref{fig:fig-6}. For
B$_{C}$, P$_{Si}$, and N$_{C}$ defects, the total magnetic moment
induced in the defective system is $0.35\mu_{B}$, $0.39\mu_{B}$,
and $0.28\mu_{B}$, respectively (refer to Table \ref{tab:table-2}),
while for $Al_{Si}$, $C_{Si}$ and $Si_{C}$ defects the system remains
non-magnetic, which is consistent with the results reported by Bekaroglu\emph{
et al. }\citep{PhysRevB.81.075433}. In the case of single defects,
i.e., V$_{C}$ and V$_{Si}$, the magnetic moment is comparatively
higher, which is $0.93\mu_{B}$, and $1.99\mu_{B}$, respectively.
In Table \ref{tab:table-2}, we have listed the resulting net magnetic
moment corresponding to various defective systems.

The complex defects (i.e., antisite-vacancy and dopant-vacancy) have
a qualitative impact on the electronic structure of the system. The
PDOS curve for the $Si_{C}$-$V_{C}$ defect indicates that the defect
states hybridize with the valence and conduction bands in the system
{[}see Fig.\ref{fig:fig-7} (j){]}, leading to a reduced band gap
$(E_{g})$ of 2.34 eV, while the defect states near the Fermi level
for $C_{Si}$-$V_{Si}$ and $C_{Si}$-$V_{C}$ {[}see Figs. \ref{fig:fig-7}(k)
and \ref{fig:fig-7}(l){]} arise mainly due to the p orbitals of the
Si and C atoms, resulting in a smaller band gap $(E_{g})$ value of
0.59 and 1.34 eV, respectively. An overlapping is observed between
the p orbital of P and Si atoms for the $P_{Si}$-$V_{C}$ $(E_{g}=0.69eV)$
defect, which indicates a strong hybridization between them. The additional
impurity states due to the p orbital of the B atom are observed for
$B_{C}$-$V_{C}$ and $B_{Si}$-$V_{C}$ defects, as shown in Figs.
\ref{fig:fig-6}(o) and \ref{fig:fig-6}(p), resulting in a significant
reduction in band gap $(E_{g})$ to 1.26 and 1.49 eV, respectively.
For $N_{C}$-$V_{C}$ $(E_{g}=1.13eV)$ and $Al_{Si}$-$V_{C}$ $(E_{g}=1.32eV)$,
the defect states arise mainly due to the p orbitals of the Si atoms,
with a very small contribution from the p orbitals of the C atom,
as is evident from Figs. \ref{fig:fig-7}(q) and \ref{fig:fig-7}(s).
The additional impurity electronic states due to the p orbital of
the C atom dominate on both sides of the Fermi level for $P_{Si}$-$V_{Si}$,
$N_{C}$-$V_{Si}$ and $Al_{Si}$-$V_{Si}$ type defects {[}see Figs.
\ref{fig:fig-7}(n), \ref{fig:fig-7}(r) and \ref{fig:fig-7}(t){]}
with a reduced band gap value of 0.39, 0.14, and 0.98 eV, respectively.
For all the cases of complex defects considered in this article, the
semiconducting nature of the material remains intact. 

We observed an interesting nonmetal-to-metal transition for $N_{C}$
and $Al_{Si}$ defects. It is important to investigate if the metallic
nature of $N_{C}$ and $Al_{Si}$ defective systems is sustained under
different defect densities. Therefore we have performed density-of-states
calculations employing the HSE06 functional using different supercell
sizes, i.e., (3 \texttimes{} 3), (5 \texttimes{} 5), and (6 \texttimes{}
6), and compared the results with a (4 \texttimes{} 4) supercell.
For the $N_{C}$ $(Al_{Si})$ defect, the Fermi level crosses the
conduction (valence) band for all supercell sizes, indicating the
metallic nature of these materials remains intact in the relatively
dense and dilute limits of such defects. We found that the conduction
and valence band edges for $N_{C}$ and $Al_{Si}$ defects are adequately
converged with different supercell sizes, as shown in Fig. S2 \citep{SupplementalMaterial},
implying a (4 \texttimes{} 4) supercell  is an appropriate choice
to study such defects in the SiC monolayer.

\subsection{Effective mass of charge carriers}

The effective masses of electrons and holes in semiconductors are
important because they determine the mobilities of the charge carriers,
subsequently influencing their photoelectric and thermoelectric properties
\citep{wang2019charge}. In this section we present our results on
the effective masses of charge carriers because, to the best of our
knowledge, no such prior calculations on the pristine and defective
SiC monolayers exist. We have computed the effective mass of electrons
and holes using the parabolic approximation around the band extremas
according to Eq. (\ref{eq:effective-mass}) \citep{feng2014effective}:

\begin{equation}
m_{k}^{*}=\hslash\left(\frac{d^{2}E(k)}{dk^{2}}\right)_{k=K_{0}}^{-1}m_{o},\label{eq:effective-mass}
\end{equation}

where $m_{o}$ is the rest mass of the electron, $\hslash$ is the
reduced Planck constant, $k$ is a wave vector in the first Brillouin
zone, $E(k)$ is the corresponding Kohn-Sham eigenvalue, and $K_{0}$
denotes the location of the band extremum. The calculated value of
electron and hole effective masses for pristine SiC monolayer at its
band extrema, i.e., $m_{M}^{e}$ and $m_{K}^{h}$, are found to be
$0.32m_{o}$ and $0.48m_{o}$, respectively. However, interestingly,
it can be noticed that the electron effective mass along the high-symmetry
directions, i.e., $m_{M\rightarrow K}^{e}$ and $m_{M\rightarrow\Gamma}^{e}$,
are found to be $3.48m_{o}$ and $0.26m_{o}$, respectively. This
is due to the decrease in the band dispersion along the $M\rightarrow K$
direction, which results in a higher effective mass. Furthermore,
the hole effective mass is found to be $1.38m_{o}$ $(m_{M\rightarrow K}^{h})$
and $0.56m_{o}$ $(m_{M\rightarrow\Gamma}^{h})$, respectively. These
results suggest that the charge carriers are more mobile along the
$M\rightarrow\Gamma$ direction as compared to the $M\rightarrow K$
direction. 

\begin{table}
\begin{tabular}{|c|c|c|c|}
\hline 
Defect-induced system & Electron effective mass $(m_{o})$ & Hole effective mass $(m_{o})$ & Total magnetic moment $(\mu_{B})$\tabularnewline
\hline 
\hline 
Pristine-SiC-monolayer & $0.32$ & $0.48$ & $0.00$\tabularnewline
\hline 
B$_{C}$ & $1.85$ & $0.60$ & $0.35$\tabularnewline
\hline 
P$_{Si}$ & $0.65$ & $0.96$ & $0.39$\tabularnewline
\hline 
N$_{C}$ & $1.59$ & $0.61$ & $0.28$\tabularnewline
\hline 
Al$_{Si}$ & $1.17$ & $0.50$ & $0.00$\tabularnewline
\hline 
C$_{Si}$ & $0.57$ & $0.45$ & $0.00$\tabularnewline
\hline 
Si$_{C}$ & $0.45$ & $0.75$ & $0.00$\tabularnewline
\hline 
V$_{C}$ & $1.07$ & $0.57$ & $0.93$\tabularnewline
\hline 
V$_{Si}$ & $1.24$ & $1.93$ & $1.99$\tabularnewline
\hline 
\emph{V}$_{Si}$-\emph{V}$_{C}$ & $0.14$ & $0.73$ & $0.00$\tabularnewline
\hline 
Si$_{C}$-\emph{V}$_{C}$ & $0.77$ & $2.04$ & $0.00$\tabularnewline
\hline 
C$_{Si}$-\emph{V}$_{Si}$ & $0.92$ & $0.45$ & $0.95$\tabularnewline
\hline 
C$_{Si}$-\emph{V}$_{C}$ & $3.90$ & $1.91$ & $0.97$\tabularnewline
\hline 
P$_{Si}$-\emph{V}$_{C}$ & $3.31$ & $0.49$ & $0.29$\tabularnewline
\hline 
P$_{Si}$-\emph{V}$_{Si}$ & $1.36$ & $3.20$ & $1.56$\tabularnewline
\hline 
B$_{C}$-\emph{V}$_{C}$ & $0.62$ & $0.66$ & $0.50$\tabularnewline
\hline 
B$_{Si}$-\emph{V}$_{C}$ & $3.29$ & $6.87$ & $0.42$\tabularnewline
\hline 
N$_{C}$-\emph{V}$_{C}$ & $0.48$ & $0.73$ & $1.30$\tabularnewline
\hline 
N$_{C}$-\emph{V}$_{Si}$ & $0.23$ & $0.25$ & $1.51$\tabularnewline
\hline 
Al$_{Si}$-\emph{V}$_{C}$ & $0.18$ & $0.84$ & $0.52$\tabularnewline
\hline 
Al$_{Si}$-\emph{V}$_{Si}$ & $1.13$ & $1.39$ & $0.52$\tabularnewline
\hline 
\end{tabular}

\caption{\label{tab:table-2}The electron and hole effective masses calculated
at their band extrema and the total magnetic moment of the pristine
and defective 2D SiC structures ($\mu_{B}$and $m_{o}$ represent
the Bohr magneton and rest mass of an electron, respectively).}
\end{table}

To understand the influence of defects on the mobilities of charge
carriers, we have carried out the effective mass calculation for the
defected systems as well. The calculated values of the electron and
hole effective masses at the band extremas for the pristine, as well
as various defected systems under the scope of this work, are listed
in Table \ref{tab:table-2}. For $B_{C}$ $(P_{Si})$ defective systems,
the electron and hole effective masses of $1.85m_{o}$ $(0.65m_{o})$
and $0.60m_{o}$ $(0.96m_{o}),$ respectively, were obtained at the
conduction (valence) band extrema. For the $C_{Si}$ $(Si_{C})$ defect
case, the electron effective mass is calculated to be $0.57m_{o}$
$(0.45m_{o}),$ while the hole effective mass is found to be $0.96m_{o}$
$(0.48m_{o}).$ The electron effective mass for $V_{Si}-V_{C}$ double
vacancy is $0.14m_{o}$, which is substantially lower than the value
$1.07m_{o}(1.24m_{o})$ of the single vacancy $V_{C}$ $(V_{Si})$
defect. On the other hand, the value of the hole effective mass for
the $V_{Si}$-$V_{C}$ defect $(0.73m_{o})$ is higher than that of
the $V_{C}$ defect $(0.57m_{o})$. Exceptionally low electron effective
mass values of $0.14m_{o}$ and $0.18m_{o}$ are obtained for the
$V_{Si}$-$V_{C}$ and $Al_{Si}$-$V_{C}$ defects, respectively,
which are comparable to the bulk silicon electron effective mass,
estimated to be $\sim0.16m_{o}$ \citep{PhysRevB.66.165217,Riffe:02}.
A low effective mass leads to a pronounced carrier transport property,
indicating the high potential of such materials in photovoltaic applications
\citep{zhang2014low,yue2018enhanced}. For the double defects C$_{Si}$-\emph{V}$_{C}$,
P$_{Si}$-\emph{V}$_{C}$, and B$_{Si}$-\emph{V}$_{C}$, the electron
effective masses are much higher than $m_{0}$, suggesting their applications
in thermoelectric devices aimed at converting otherwise wasted heat
into electric power \citep{bhattacharya2014role}. 

\subsection{Influence of defects on the optical properties }

\begin{figure}
\includegraphics[scale=0.55]{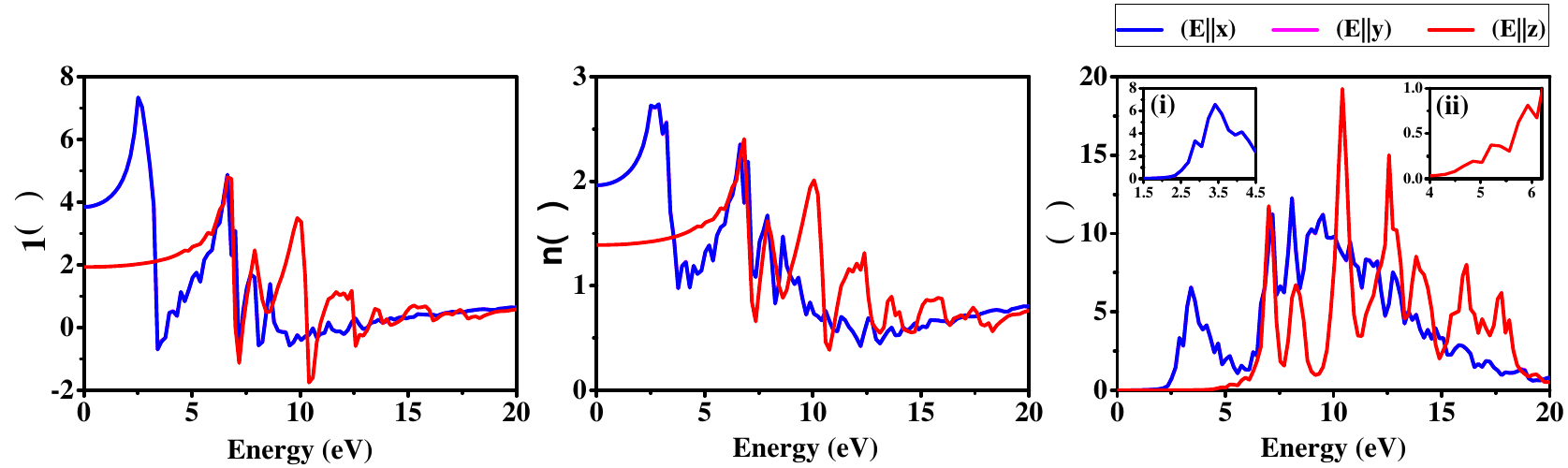}\caption{\label{fig:fig-8}(a) The real part of dielectric $\varepsilon_{1}(\omega)$,
(b) refractive index $n(\omega)$ , and (c) absorption coefficient
$\alpha(\omega)$, plotted for pristine SiC monolayer}
\end{figure}

In this section we present and discuss our results on the optical
absorption spectra of the pristine and defective SiC monolayers with
the aim of understanding the influence of defects on their optical
properties. Our findings give insights into the rich electronic and
optical properties of this newly synthesized SiC monolayer for its
potential application in electronic and optoelectronic devices of
the next generation. For this purpose we have computed the frequency-dependent
real $\varepsilon_{1}(\omega)$ and imaginary $\varepsilon_{2}(\omega)$
part of the dielectric functions using the orbitals and the band structures
obtained from the GGA-PBE calculations. The imaginary part was computed
by performing the sum over the conduction-band states, while the real
part was obtained from the imaginary part using the Kramers--Kronig
(KK) relationship. Thus, these calculations were performed within
the independent-electron model, ignoring the excitonic effects that
arise due to the electron-hole interactions. To include the influence
of excitonic effects, one must perform calculations using approaches
such as the time-dependent density-functional theory (TDDFT) or the
Bethe-Salpeter equation (BSE) that take into account the quantum many-particle
effects. However, the purpose behind the present calculations is to
obtain a semiquantitative understanding of the influence of defects
on the optical properties, for which we feel that the independent-electron
approach is sufficient. The following equations (\ref{eq:refractive-index}
and \ref{eq:optical-absorption}) are used to compute the relevant
refractive index and optical absorption of the system with respect
to the photon energy. 

\begin{equation}
n(\omega)=\left[\frac{\sqrt{\varepsilon_{1}^{2}(\omega)+\varepsilon_{2}^{2}(\omega)}+\varepsilon_{1}(\omega)}{2}\right]^{\frac{1}{2}},\label{eq:refractive-index}
\end{equation}

\begin{equation}
\alpha(\omega)=\frac{2\omega}{c}\left[\frac{\sqrt{\varepsilon_{1}^{2}(\omega)+\varepsilon_{2}^{2}(\omega)}-\varepsilon_{1}(\omega)}{2}\right]^{\frac{1}{2}},\label{eq:optical-absorption}
\end{equation}
 For the pristine SiC monolayer, absorption starts at around 2.23
eV, with its first absorption peak at 2.87 eV, followed by a second
absorption peak at 3.41 eV for the light polarized in the $xy$ plane
($\parallel$-polarized), i.e., with the electric field of the radiation
$E\parallel x$ or $y$ directions {[}\ref{fig:fig-8}(c){]}. We also
note that for the $\parallel$-polarized light, the optical response
of the pristine monolayer is fully isotropic, i.e., the same for both
$x$ and $y$ polarization directions. While for the $z$-polarized
light ( $E\parallel z$, or $\perp$-polarized), the absorption starts
at a higher energy of 4.30 eV, followed by the first absorption peak
at 4.81 eV, which is expected for two-dimensional materials since
more energy is required to excite electrons in a direction perpendicular
to the plane of the material because of the stronger confinement in
that direction. Several other peaks are located in the ultraviolet
region of the spectrum, indicating a number of transitions from the
valence band to the conduction band. The dominant peak corresponding
to the maximum absorption is observed at 8.09 eV ($\parallel$-polarized)
and 10.41 eV ( $\perp$-polarized). Table \ref{tab:table-3} shows
the first and the most intense absorption peak values for the pristine
SiC monolayer and the bands that contribute to the single electron
transition corresponding to the peak. In order to determine the conduction
bands involved in these transitions, we compared the excitation energies
of the first and the most intense peaks of the absorption spectra
with the energy difference between highest valence band $(v)$ and
the $n^{th}$ conduction band $(c+n)$, calculated using the PBE-GGA
band structures {[}see Fig. S3 \citep{SupplementalMaterial}{]}.

We found that the transition occurs at the $\Gamma$ point for the
first and most intense peaks for the $\perp$-polarized light, while
the transition is at K and $\Gamma$ points, respectively, for the
first, and the most intense peaks in case of $\parallel$-polarized
light. 

\begin{table}
\begin{tabular}{|c|c|c|}
\hline 
\multirow{2}{*}{Peak} & \multicolumn{2}{c|}{Peak locations (eV)}\tabularnewline
\cline{2-3} \cline{3-3} 
 & $\parallel$-polarized & $\perp$-polarized\tabularnewline
\hline 
First peak & 2.87 $(\nu\rightarrow c)$ & 4.81 $(\nu\rightarrow c+14)$\tabularnewline
\hline 
Most intense peak & 8.09 $(\nu\rightarrow c+41)$ & 10.41 $(\nu\rightarrow c+59)$\tabularnewline
\hline 
\end{tabular}

\caption{\label{tab:table-3}Calculated first and most intense absorption peak
energy values for the pristine SiC monolayer from optical absorption
spectra. $\nu$ and $c$ correspond to the VBM and CBM, while $c+n$
corresponds to the $n^{th}$ conduction band above the lowest conduction
band $c$.}

\end{table}

The static dielectric constant $\varepsilon_{1}(0)$ and the refractive
index $n(0)$ are found to be 3.85 (1.93) and 1.96 (1.39) for $\parallel$-polarized
($\perp$-polarized) light, as shown in Figs. \ref{fig:fig-8}(a)
and \ref{fig:fig-8}(b). For the photon energy ranges 3.6-5.6 eV and
8.1 eV and higher for the $\parallel$-polarized light, the refractive
index ($n$) of SiC monolayer is found to be less than that of the
glass ($<n_{glass}$), indicating comparatively higher transmission
and hence low absorption. While for the$\perp$-polarized light, higher
transmission will be observed for the energy range 0.0-4.3 eV.

\begin{figure}
\includegraphics[scale=0.55]{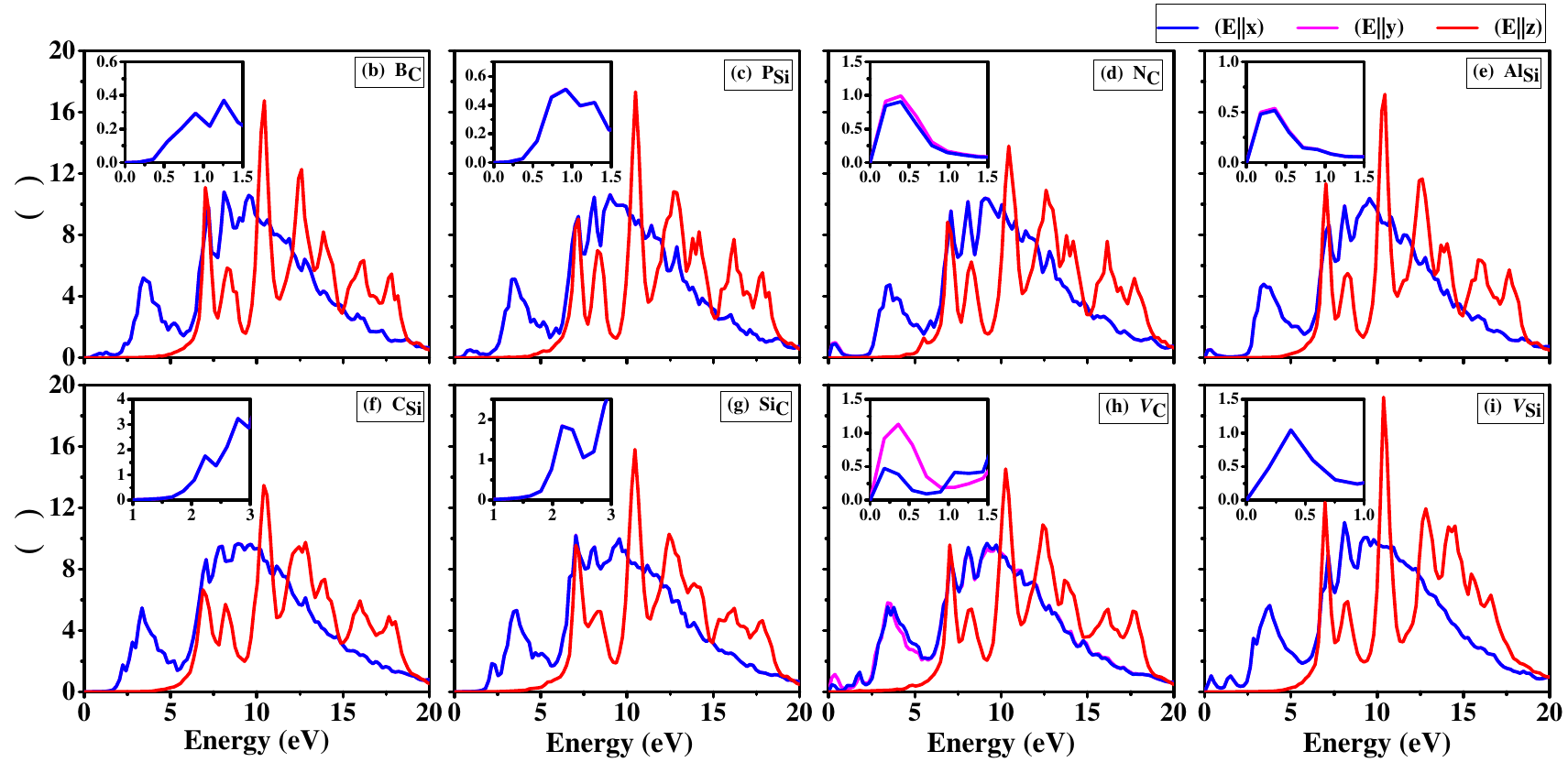}\caption{Optical absorption spectra of the SiC monolayer with simple defects
: (a) $B_{C}$, (b) $P_{Si}$, (c) $N_{C}$, (d) $Al_{Si}$, (e) $C_{Si}$,
(f) $Si_{C}$, (g) $V_{C}$, and (h) $V_{Si}$ \label{fig:fig-9}}
\end{figure}

\begin{figure}
\includegraphics[scale=0.55]{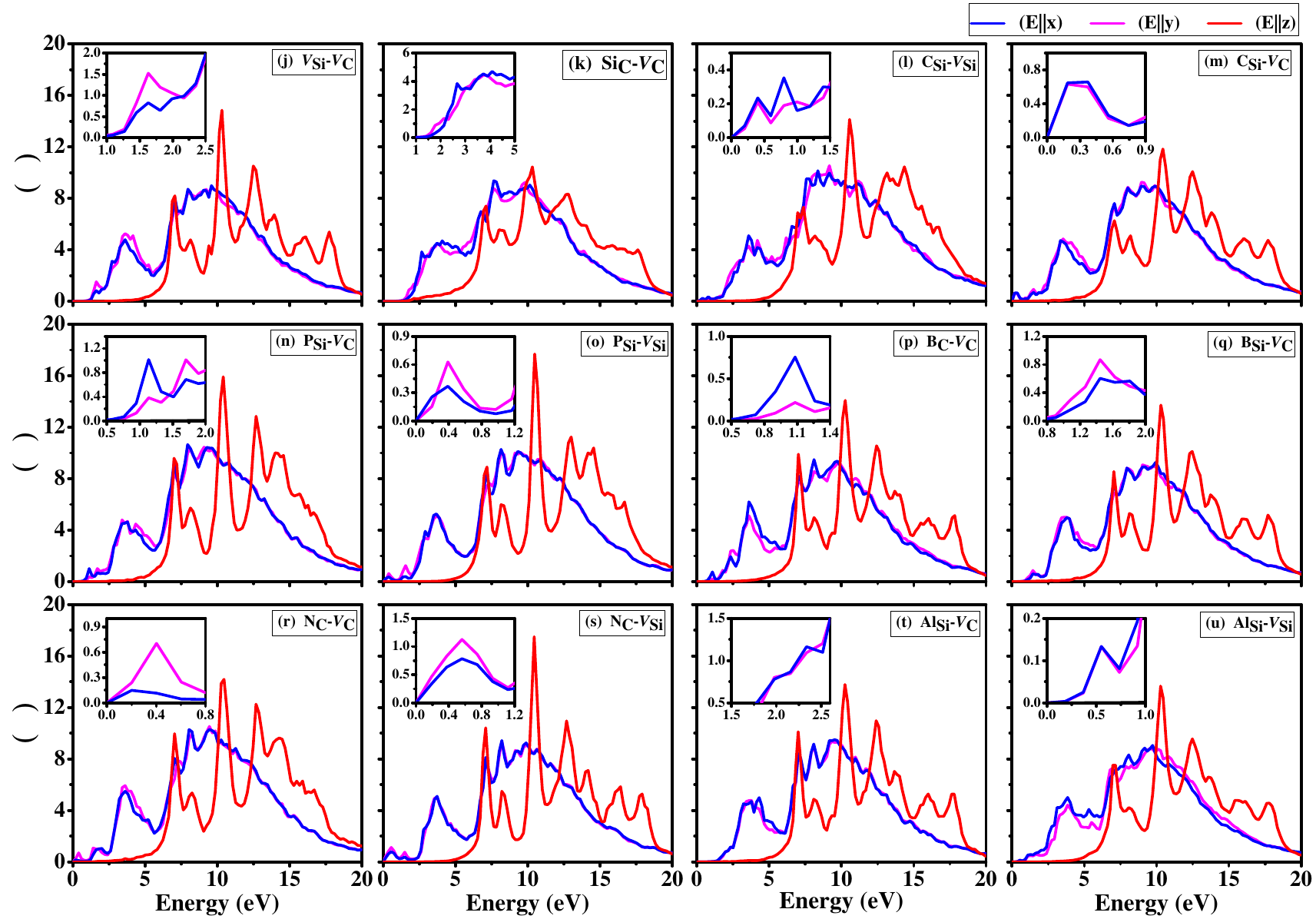}

\caption{Optical absorption spectra of the SiC monolayer with complex defects
: (a) $V_{Si}$-$V_{C}$, (b) $Si_{C}$-$V_{C}$, (c) $C_{Si}$-$V_{Si}$,
(d) $C_{Si}$-$V_{C}$, (e) $P_{Si}$-$V_{C}$, (f) $P_{Si}$-$V_{Si}$,
(g) $B_{C}$-$V_{C}$, (h) $B_{Si}$-$V_{C}$, (i) $N_{C}$-$V_{C}$,
(j) $N_{C}$-$V_{Si}$, (k) $Al_{Si}$-$V_{C}$, and (l) $Al_{Si}$-$V_{Si}$
\label{fig:fig-10}}

\end{figure}

As can be seen from calculated optical absorption data (Figs. \ref{fig:fig-9}
and \ref{fig:fig-10}), the optical absorption for the defective SiC
monolayers starts at much lower energies, as compared to the pristine
one. The reason behind the onset of absorption at lower energies in
the defective systems is the presence of defect levels in the midgap
region that effectively reduce the optical gap in these materials.
For the $\parallel$-polarized light, the absorption starts at around
0.36 eV for $B_{C}$ and $P_{Si}$ defects due to the additional impurity
states generated above and below the Fermi level. The first absorption
peak is observed in the infrared (IR) region of the spectrum at 0.89
eV and 0.74 eV for $B_{C}$ and $P_{Si}$ defects, respectively. For
$N_{C}$ and $Al_{Si}$, the defect states appear at the Fermi level
(Fig. \ref{fig:fig-7}) and hence the absorption starts at 0 eV, indicative
of the metallic nature of these systems. The first peak is found at
0.20 eV for $N_{C}$ and $Al_{Si}$ defects. Similarly, due to the
impurity states being very close to the Fermi level, the onset of
absorption starts again near zero photon energy for $V_{C}$ and $V_{Si}$
defects, with the first absorption peak at 0.17 and 0.38 eV, respectively.
A number of defect states appear above and below the Fermi level,
which leads to more than one absorption peak within a small energy
interval in the infrared region of the spectrum. It could be noted
that a slight anisotropy in the optical absorption in the xx and yy
direction is observed for the $V_{C}$ defect. For the cases of $C_{Si}$
and $Si_{C}$, the defect states appear close to CBM and VBM, respectively.
Thus the absorption spectrum does not change significantly as compared
to the pristine monolayer, with the first absorption peaks at 2.24
and 2.17 eV, respectively. Furthermore, no significant changes are
observed in the optical absorption spectrum for the $\perp$-polarized
light for the systems with simple defects as compared to the pristine
monolayer.

Next we discuss the absorption spectra of monolayers with complex
defects. For the $\parallel$-polarized light, the absorption starts
at around 1.17 eV leading to the first absorption peak at 1.63 eV
for the $V_{Si}$-$V_{C}$ defect. For $C_{Si}$-$V_{Si}$, $P_{Si}$-$V_{Si}$,
and $N_{C}$-$V_{Si}$ defects, the absorption starts near 0 eV, owing
to their very low band gap. Their first absorption peak is observed
at 0.40, 0.36, and 0.54 eV, respectively. The absorption spectra of
the $Si_{C}$-$V_{C}$ defect is similar to the pristine case with
the first absorption peak shifted to 2.67 eV. A sharp absorption peak
occurs at 1.13 and 1.10 eV for $P_{Si}$-$V_{C}$ and $B_{C}$-$V_{C}$
defects. Broad absorption peaks at 1.44, 1.72, and 0.18 eV corresponding
to $B_{Si}$-$V_{C}$, $N_{C}$-$V_{C}$, and $C_{Si}$-$V_{C}$ defects,
respectively, indicate several electronic transitions from occupied
energy states to unoccupied energy states within that energy interval.
$Al_{Si}$-$V_{C}$ and $Al_{Si}$-$V_{Si}$ have new absorption peaks
at $\sim$ 1.99 and 0.54 eV, respectively, while the second absorption
peak splits into more than one peak, indicating multiple transitions.
Notably, a small anisotropy in the optical absorption is observed
for the $\parallel$-polarized light with respect to $x$ and $y$
polarization directions for all the complex defects under consideration.
For the $\perp$-polarized light, no additional absorption peaks or
shift of absorption peaks is observed, except that a significant decrease
(0.54 times) in the peak strength is noticed for the $Si_{C}$-$V_{C}$
defect. 

As is known, materials which show absorption in the visible region
are suitable for photocatalysis using sunlight. Thus these defective
systems can be utilized for photocatalysis. Several defective systems
showed absorption peaks in the infrared (IR) region of the spectrum,
which makes them crucial for the rational design of IR LEDs for their
application in sensing, communication devices, remote control, etc.
The optical absorption spectra of defective systems also cover a wide
range of visible and UV regions, indicating that these materials can
have important applications in optoelectronic devices in a broad frequency
range.

\section{CONCLUSION\label{sec:conclusion}}

In this paper we presented a systematic and detailed first-principles
DFT-based study of the stability, electronic structure, and optical
properties of both pristine and defective SiC monolayers. A variety
of single as well as double atom defects were considered employing
$4\times4$ supercells, and their influence on a number of properties,
such as the band gap, effective electron and hole masses, and optical
absorption spectra, were studied. The negative adhesion energy of
SiC monolayer on C and Ta-terminated TaC \{111\} film indicates that
SiC monolayer can be stabilized upon metal carbide film. Furthermore,
by calculating the formation energies of all the defective configurations,
within both the Si and C rich conditions, we demonstrated that the
defective structures can be easily synthesized. 

To summarize, we found that the electronic structure and optical properties
of the SiC monolayer can be tuned in a variety of ways by introducing
defects in a controlled manner. Analyzing the calculated theoretical
formation energies of various defects, we predict that $N_{C}$ defect
could be naturally present in silicon carbide monolayer even during
growth due to its very low formation energy. The formation energy
related to Si-vacancy defects is found to be higher as compared to
C-vacancy-related defects. This behavior is expected since the Si-atom
has a greater atomic mass than the C-atom and will be difficult to
remove. We also verified that the trend in formation energy remains
the same with different defect concentrations. The structural solidity
at room temperature for defected systems has been verified by employing
\emph{ab-intio} molecular dynamics (AIMD) simulations. The in-depth
analysis of the electronic band structures and the atom projected
density of states (PDOS) showed that the additional impurity states
are generated within the forbidden region of pristine silicon carbide
in the presence of defects. These impurity states lead to the dramatic
reduction in the band gap, which can improve the electrical conductivity
of the material. An interesting transition from semiconducting to
metallic is observed for $N_{C}$ and $Al_{Si}$ defective systems.
Furthermore, we also calculated the electron and hole effective masses,
which is one of the fundamental characteristics of the semiconductor.
We observed that the electron effective mass for pristine silicon
carbide monolayer is significantly higher $(\sim3.48m_{o})$ along
the $M\rightarrow K$ direction, which will lead to better thermoelectric
performance. Interestingly, the electron effective mass at the band
extrema for $V_{Si}$-$V_{C}$ and $Al_{Si}$-$V_{C}$ defects is
found to be very low, which is comparable to the bulk silicon effective
mass. Such low-effective mass materials are suitable for high-speed
electronic devices. The real and imaginary parts of the frequency-dependent
dielectric function have been used to calculate the refractive index
and optical absorption coefficient. In terms of the onset of optical
absorption, both the pristine and defective monolayers exhibit significant
anisotropy with respect to the light polarized parallel to the plane
of the monolayer as compared to the perpendicularly-polarized light.
This can be seen as a structural fingerprint in the optical response,
which can be used for the detection of the monolayers employing optical
spectroscopy. Further, it has been found that the absorption edge
shifts towards the lower energy range of the spectrum for defect-induced
systems as compared to its pristine counterpart. The additional absorption
peak is observed at 0.89 eV and 0.74 eV for $B_{C}$ and $P_{Si}$
defects ($\parallel$-polarized light), respectively. While for $C_{Si}$
and $Si_{C}$ defects, the first absorption peak shifts to 2.24 and
2.17 eV ($\parallel$-polarized light), respectively, since the defect
states are generated very close to the conduction band and valence
band, respectively. Therefore, these nanosystems exhibit excellent
ability of optical absorption in the near-infrared and visible regions
of the spectrum, making them viable for future photovoltaic and optical
devices. Our results will help to speed up the understanding of defects
in 2D SiC monolayer to be realized experimentally in the near future.

\newpage{}

\bibliographystyle{apsrev4-2}
\addcontentsline{toc}{section}{\refname}\bibliography{references}

\end{document}


\title{Supporting Information (SI): Defect-driven tunable electronic and
optical properties of two-dimensional silicon carbide}
\author{Arushi Singh}
\email{arushi.phy@iitb.ac.in}

\affiliation{Department of Physics, Indian Institute of Technology Bombay, Powai,
Mumbai 400076, India}
\author{Vikram Mahamiya}
\email{mahamiyavikram@gmail.com}

\affiliation{National Institute for Materials Science (NIMS), 1-1 Namiki, Tsukuba,
Ibaraki 3050044, Japan}
\affiliation{Department of Physics, Karpagam Academy of Higher Education, Coimbatore
641021, Tamil Nadu India}
\affiliation{Centre for Computational Physics, Karpagam Academy of Higher Education,
Coimbatore 641021, Tamil Nadu, India}
\author{Alok Shukla}
\email{shukla@iitb.ac.in}

\affiliation{Department of Physics, Indian Institute of Technology Bombay, Powai,
Mumbai 400076, India}

\maketitle
\begin{figure}
\includegraphics[scale=0.6]{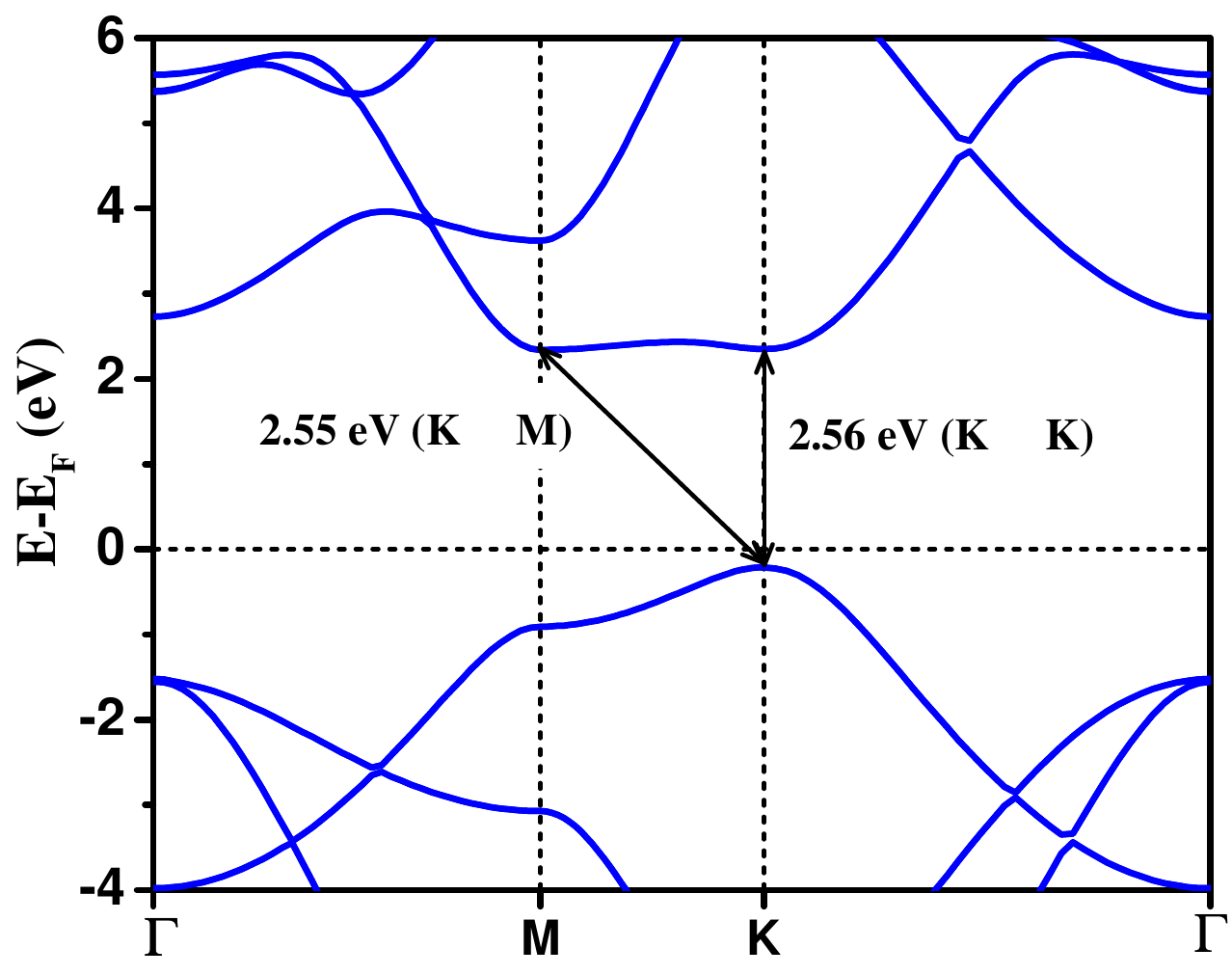}

\caption{\label{fig:S1}Electronic band structure of a primitive cell of 2D-SiC
monolayer calculated using PBE-GGA approach }

\end{figure}

Figure \ref{fig:S1} demonstrates the PBE-functional-based band structure
of SiC monolayer. The conduction edge at the K and M points of the
of the Brillouin zone are almost at the same energy level. Therefore,
we calculated the PBE band structure employing strict convergence
criteria, i.e., using higher energy cut-off (700 eV) and denser K-grid
(19\texttimes 19\texttimes 1) and we found an indirect band gap of
2.55 eV and a direct band gap of 2.56 eV. 

\begin{figure}
\includegraphics[scale=0.6]{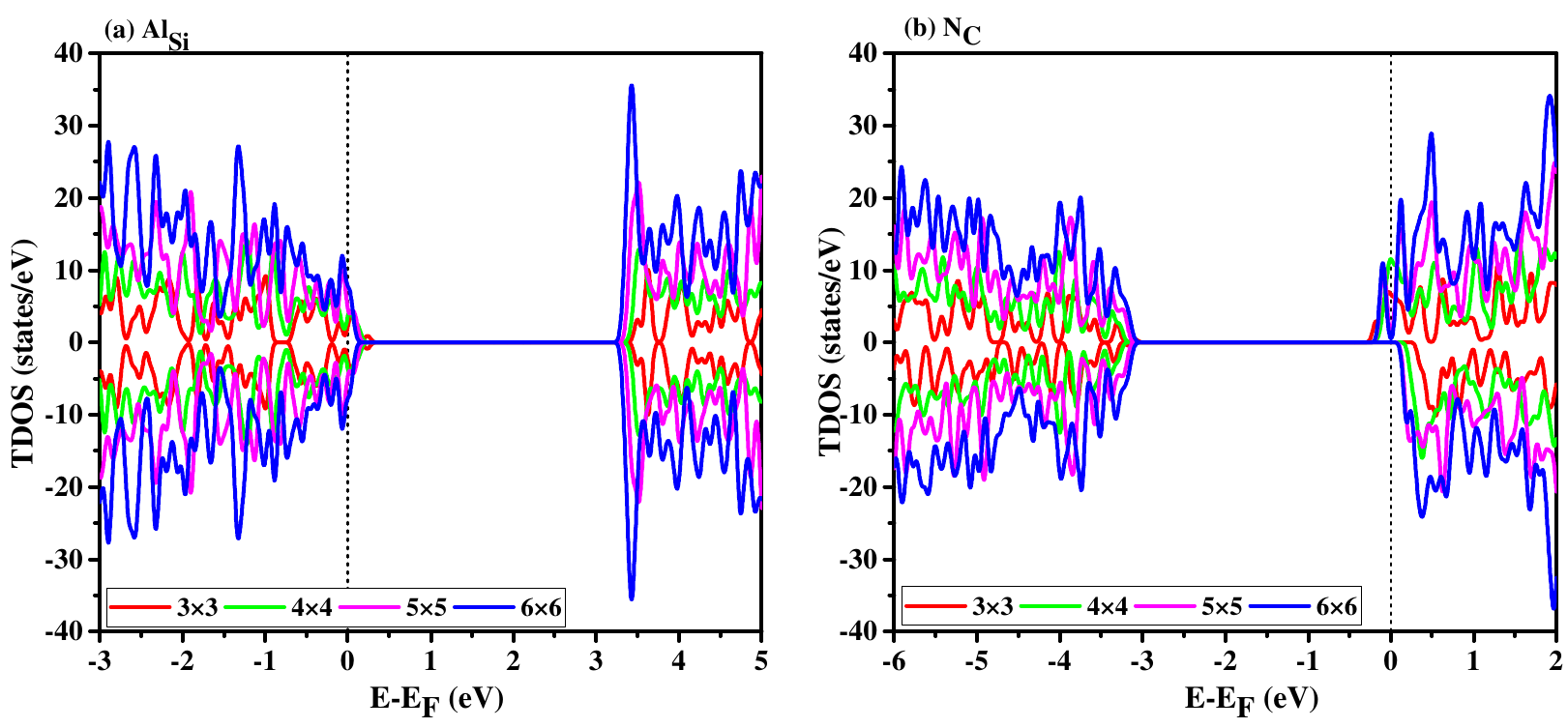}

\caption{Calculated HSE06 TDOS (total density of states) for (a) $Al_{Si}$,
and (b) $N_{C}$ defects using different supercell sizes i.e., (3\texttimes 3),
(4\texttimes 4), (5\texttimes 5) and (6\texttimes 6) .}
\end{figure}

\begin{figure}

\includegraphics[scale=0.6]{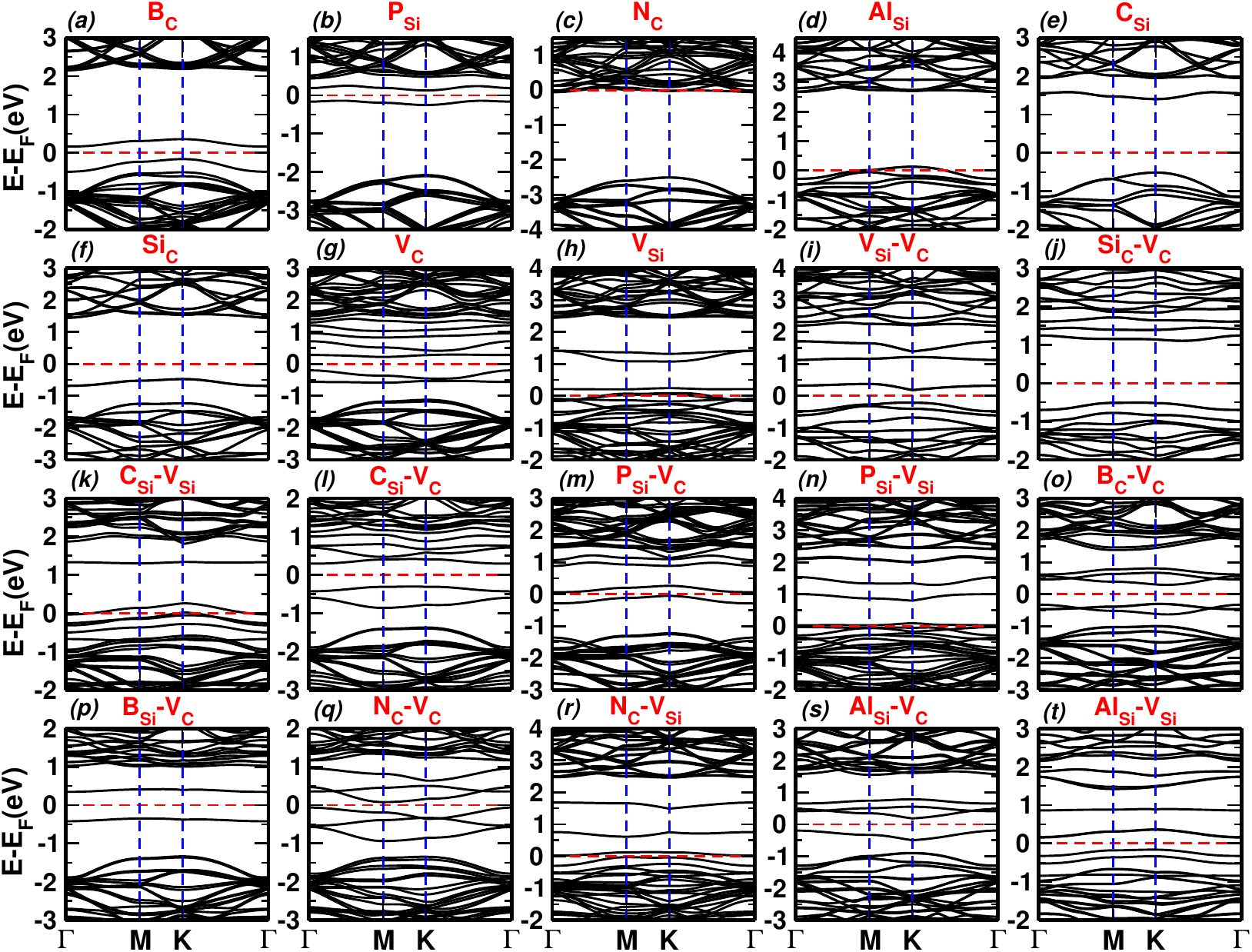}\caption{Electronic band structures of various defects under investigation
computed using the PBE-GGA approach: (a) $B_{C}$, (b) $P_{Si}$,
(c) $N_{C}$, (d) $Al_{Si}$, (e) $C_{Si}$, (f) $Si_{C}$, (g) $V_{C}$,
 (h) $V_{Si}$, (i) $V_{Si}$-$V_{C}$, (j) $Si_{C}$-$V_{C}$, (k)
$C_{Si}$-$V_{Si}$, (l) $C_{Si}$-$V_{C}$, (m) $P_{Si}$-$V_{C}$,
(n) $P_{Si}$-$V_{Si}$, (o) $B_{C}$-$V_{C}$, (p) $B_{Si}$-$V_{C}$,
(q) $N_{C}$-$V_{C}$, (r) $N_{C}$-$V_{Si}$, (s) $Al_{Si}$-$V_{C}$,
and (t) $Al_{Si}$-$V_{Si}$ }
\end{figure}